\documentclass[letterpaper,twocolumn,10pt]{article}
\usepackage{usenix2019_v3}

\usepackage{tikz}
\usepackage{amsmath}

\usepackage{filecontents}

\usepackage{changepage}
\usepackage{xcolor}

\usepackage{enumerate}

\usepackage{array}
\newcommand{\PreserveBackslash}[1]{\let\temp=\\#1\let\\=\temp}
\newcolumntype{C}[1]{>{\PreserveBackslash\centering}p{#1}}
 \newcolumntype{R}[1]{>{\PreserveBackslash\raggedleft}p{#1}}
\newcolumntype{L}[1]{>{\PreserveBackslash\raggedright}p{#1}}
\usepackage{textcomp}
\usepackage{booktabs}
\usepackage{threeparttable}
\usepackage{multirow}
\usepackage{caption}
\usepackage{subcaption}
\usepackage{amsmath,amssymb,amsthm}
\usepackage{amsmath, bm}
\usepackage{algorithm}
\usepackage{algorithmic}
\newtheorem{myDef}{Definition}
\newcommand{\keypoint}[1]{\vspace{0.1cm}\noindent\textbf{#1}}
\newcommand{\cut}[1]{}

\newcommand{\WFA}{WFA}
\newcommand{\WFD}{WFD}
\newcommand{\ST}{ST-WFA}
\newcommand{\MT}{MT-WFA}
\begin{document}

%-------------------------------------------------------------------------------

%don't want date printed
\date{}

% \title{\Large \bf Formatting Submissions for a USENIX Conference:\\
%   An (Incomplete) Example}
\title{\Large \bf End-to-End Multi-Tab Website Fingerprinting Attack: A Detection Perspective}

% \author{%
% 		Mantun Chen\addresslink{1},
%   	Yongjun Wang\addresslink{1},
%   	Zhiquan Qin\addresslink{1} and
%   	Xiatian Zhu\addresslink{2}
%     }

% \address[1]{College of Computer, National University of Defense Technology, Changsha 410073, China}
% \address[2]{University of Surrey, Stag Hill, University Campus, Guildford GU2 7XH, UK.}

% \correspondence{Correspondence should be addressed to 
%     	Yongjun Wang: wangyongjun@nudt.edu.cn}

% \email{chenmantun19@nudt.edu.cn (Mantun Chen), wangyongjun@nudt.edu.cn (Yongjun Wang), tanzhiquan14@nudt.edu.cn (Zhiquan Qin) and eddy.zhuxt@gmail.com (Xiatian Zhu)}%
%for single author (just remove % characters)
\author{
	{\rm Mantun Chen, Yongxin Chen$^{Co~first~author}$, Yongjun Wang$^{Corresponding~author}$, Peidai Xie, Shaojing Fu}\\
	College of Computer, National University of Defense Technology\\
	\{chenmantun19, yongxinchen\_cx, wangyongjun, tanzhiquan14, fushaojing\}@nudt.edu.cn\\ xpd2002@126.com
	\and
	{\rm Xiatian Zhu}\\
	University of Surrey\\
	eddy.zhuxt@gmail.com
}
% copy the following lines to add more authors
% \and
% {\rm Name}\\
%Name Institution
%} % end author

\maketitle

%-------------------------------------------------------------------------------
\begin{abstract}
%-------------------------------------------------------------------------------
% Your abstract text goes here. Just a few facts. Whet our appetites.
% Not more than 200 words, if possible, and preferably closer to 150.
Website fingerprinting attack (\WFA)  aims to deanonymize the website a user is visiting through anonymous networks channels (e.g., Tor).
Despite of remarkable progress in the past years, 
most existing methods make implicitly a couple of {\em artificial assumptions}
that (1) only a single website (i.e., single-tab) is visited each time,
and (2) website fingerprinting data are pre-trimmed into a single 
% website per trace
trace per website
manually.
In reality, a user often open multiple tabs for multiple websites
spontaneously.
% , particularly with the extra loading cost added 
% by anonymous networks channels. 
Indeed, this multi-tab \WFA~(\MT) setting has been studied 
in a few recent works, but all of them still fail to fully respect 
the real-world situations.
In particular, the overlapping challenge between website fingerprinting
has never been investigated in depth.
In this work, we redefine the problem of \MT~as {\em detecting multiple 
% {\color{blue}target website fingerprinting instances}
monitored traces, given a natural untrimmed traffic data including 
% {\color{black}target website fingerprinting, background traffic}
monitored traces, unmonitored traces, and potentially unconstrained overlapping between them}.
This eliminates the above assumptions,
going beyond the conventional {\em single website fingerprint classification} perspective taken by all previous \WFA~methods.
To tackle this realistic \MT~problem,
we formulate a novel {\bf \em Website Fingerprint Detection} (WFD)
model capable of detecting accurately the start and end points of 
% a target fingerprint instance 
{\color{black}all the monitored traces}
and classifying them jointly, given long, untrimmed raw traffic data.
WFD is end-to-end, with the 
% feature extraction and fingerprint detection 
{\color{black} trace localization and website classification}
integrated in a single unified pipeline.
To enable quantitative evaluation in our \MT~setting,
we introduce new performance metrics.
Extensive experiments on several newly constructed benchmarks
show that our WFD outperforms the state-of-the-art alternative methods
in both accuracy and efficiency by a large margin,
even with a very small training set. 
%
% We design new evaluation metrics and conduct extensive evaluations.
% We show our approach could work significantly better than the state-of-the-art method even training with a very small data amount. 
% Specifically, its attack speed is marvelous so that it can be utilized to simultaneously attack many users at real time and on a large scale. 
Code is available at
\url{https://github.com/WFDetector/WFDetection}.
\end{abstract}

%-------------------------------------------------------------------------------
\section{Introduction}
% Nowadays, Internet users pay more attention to the protection of their browsing privacy and choose anonymous networks to ensure it. 
% Tor, one of the most popular anonymous networks, has helped several million privacy-conscIoUT users a day to defend against network surveillance \cite{Tor2021}.
With the onion routing \cite{Dingledine2004}, Tor disassociates several million privacy-conscious users a day with their visiting websites by forwarding the multi-layer encrypted packets through a number of volunteer proxies \cite{Tor2021}. 
% Even with the help of Tor, the encrypted digital footprints would inevitably be exposed to a network eavesdropper tapped on any one of the on-path routers before entering Tor's entry proxy. 
However, by observing the side-channel pattern
without decryption on any one of the on-path routers before entering Tor's entry proxy, an attacker could still compromise a user's browsing privacy via traffic analysis, i.e.,
website fingerprinting attack (\WFA). 
% Since the \WFA~is virtually impossible to detect, its threat would be disastrous if Tor is vulnerable. 
In the past decade, an extensive number of existing studies \cite{Abe2016,Hayes2016,Panchenko2016,Panchenko2011,Rimmer2018,Sirinam2018,Wang2014, Wang2013, Wang2016,Bhat2019} have 
% claimed that they have 
achieved remarkable success,  
% Recently, by leveraging hand-designed feature engineering or deep learning based automatic feature extraction, many studies \cite{Abe2016,Hayes2016,Panchenko2016,Panchenko2011,Rimmer2018,Sirinam2018,Wang2014, Wang2013, Wang2016,Bhat2019} have 
% claimed that they have 
% achieved remarkable success in WF attack against Tor, 
%
even with only a few training samples per website \cite{Sirinam2019, CHEN2021cn, CHEN2021scn}.

Nonetheless, most existing methods make a couple of {\em artificial assumptions} implicitly: 
(1) 
Each time a user visits only a single website
 and visit 
%  {\color{black}While a user visits only a single website each time, he/she visits?}
a number of websites separately and sequentially, i.e., {\bf\em single-tab \WFA~(\ST)};
(2) Raw website fingerprinting data are pre-trimmed manually so that a single 
trace involves one website of interest.
These are obviously not true in typical real-world situations.
Often, a user would open multiple tabs to visit different websites
spontaneously and change their focus from one tab to another over time,
i.e., {\bf\em multi-tab WFA~(\MT)}.

% However, they have been criticized for overestimating the attacker's abilities because they are grounded on a too strong single-tab assumption \cite{Juarez2014,Gu2015, Xu2018, Wang2016}.
% In fact, only visiting one page each time is almost impossible.

% While Tor is slow, a user would not wait for the page loading in the current tab, but load a new page or open new tabs to visit other web pages, called  multi-tab assumption.
% For the multi-tab trace, it is very hard to split the consecutive traces into single ones correctly.
% Moreover, the difficult overlap situations need to be well considered.
% Compared with the hot research of \ST, there are very limited studies about \MT~\cite{Gu2015, Xu2018, Wang2016,Gong2020}. 
% Given the previous three studies \cite{Gu2015, Xu2018, Gong2020} focused on Tor's \MT, t
In the literature there are only very few studies on \MT~\cite{Gu2015, Xu2018, Wang2016,Gong2020}. 
Importantly, they are highly limited in different aspects, 
and none of them investigate {\em fully} the 
real-world user behaviors and all the native challenges with \MT.
Specifically, 
Gu et al. \cite{Gu2015} only discussed the classification of two-tab traces for SSH without considering the localization of individual traces at all. % and ignores the detection.
Wang et al. \cite{Wang2016} moved a step further by studying how to split Tor's two-tab traces so that the \MT~problem can be solved by previous \ST~models.
Later on, Xu et al. \cite{Xu2018} improved the splitting of two-tab traces but limited to only classifying the clean front segment of the first website fingerprinting.
More recently, Gong et al. \cite{Gong2020} considers
a simplified \MT~setting without overlapping between website fingerprinting in a defense perspective.

In this work, we for the first time investigate a realistic \MT~problem setting. Unlike previous attempts above, we minimize the introduction of artificial assumptions and conditions for enabling a true indication of model's deployability and scalability in the application of real-world \WFA~tasks.
Concretely, we redefine the problem of \MT~as follows:
{The {\em input} is raw untrimmed traffic data
% with 
{\color{black}of}
multi-tab traces potentially with
% and 
% {\color{black}in which there exists?}
random overlapping between adjacent visits. The {\em objective} is to find where (start and end) and what (website label) each individual full monitored trace 
of 
% {\color{black}from?}
any monitored website is.
% and  the location and label of all the single full traces of interest should be detected without manual trimming.
}
% (1) The input is raw traffic data without manual trimming 

% (2) Multiple traces of interest in a single raw traffic data

% (3) Overlapping between website fingerprinting

% {\color{black} \bf Tun to help above}

% only studied the split and classification of the consecutive multi-tab traces without consideration of the overlap.
% They have only scratched the surface of \MT~problem.
% Gu et al. \cite{Gu2015} only discussed the classification of SSH's two-tab traces and ignores the detection.
% Wang et al. \cite{Wang2016} studied the split of Tor's two-tab traces so that the \MT~problem can be transformed into the \ST~problem.
% Xu et al. \cite{Xu2018} tried to split the Tor's two-tab trace more accurately and only focused on the first page classification with its clean front segment.
% Gong et al. \cite{Gong2020} only studied the split and classification of the consecutive multi-tab traces without consideration of the overlap.
% For \MT, they all did not consider the location but just the classification.

To solve the proposed \MT~problem,
we introduce the first end-to-end
{\bf \em Website Fingerprint Detection} (WFD)
model.
This is a result of our drawing inspiration from the success of
object detection methods in computer vision,
viewing a 
{\color{black}single full monitored trace}
% website fingerprinting 
as an object
and  
{\color{black}a long multi-tab trace}
% a raw traffic trace 
as an image.
Along with feature extraction, localization and classification of all full monitored traces are jointly optimized in a single deep learning architecture, i.e., end-to-end.
This is drastically different from and more elegant than all the previous \WFA~methods \cite{Gong2020,Xu2018} that model the localization (i.e., trace splitting) and classification {\em independently} and hence
suffer from the notorious localization error propagation problem.
As a consequence, their methods are poor in 
tackling the realistic WFA situations (see Table \ref{tab:overlap_all}).

% From the perspective of method, their shared characteristics are the solution of split and classification. 
% Their weakness is that there exists a paradox between the two stages.
% Split means the attacker wants to locate the position of every single trace: where to start and where to end. 
% This demands the attacker already knows how to distinguish  different trace segments in the multi-tab trace from each other before splitting.
% In fact, a perfect distinction equals an ideal  classification.
% Since a well distinction have been done, then the classification stage becomes meaningless. 
% If we do classification first, the splitted clean segment must be prepared firstly because the performance of existing \ST s would drop dramatically given mixed trace.
% Hence, the above two-stage solutions cannot work well.

% While rethinking the \MT~problem, we find that 
% only the positions and class labels of trace segments from monitored websites are all we need to attack, and the other segments from unmonitored are not important.
% Regarding that finding monitored trace segments in the multi-tab trace is similar to detecting some object in one figure or someone's voice in the noisy party, we think the \MT~should focus on
% detecting the monitored trace segments hiding in the background of valueless segments such as unmonitored ones.
%

Our {\bf contributions} are summarized as follows:
\begin{description}
\item{\bf (I)} We investigate for the first time the realistic and more challenging \MT~problem. Instead of classifying individual, manually trimmed trace segments as in existing methods, we take a {\em detection} perspective. 
This eliminates the unrealistic assumptions of sequential single website visiting and manual trimming. 
For quantitative evaluation, we introduce new performance metrics (mean Average {\color{black}Precision (mAP), and MB per second (MBps))} for \MT.
%
%Aiming to find and recognize monitored trace segments, we first reformulate the \MT~problem as a WF detection problem.
\item{\bf (II)}
We propose the first end-to-end {\bf \em Website Fingerprint Detection} (WFD) model.
This is based on our perspective of considering each monitored trace of interest as a specific object in an image.
Hence this allows us to adapt the success of object detection in {\color{black}computer vision} for \MT.
In design, our model is superior
as all the model components (feature extraction, 
{\color{black}scale encoding/perception, and two-head prediction})
% localization, and classification
of WFD can be optimized jointly in training,
which is impossible with existing alternative methods.
To improve the model efficiency, 
we introduce a novel compact input representation (namely Burst),
{\color{black}specially-designed lightweight components, and a fast training strategy.}
% and specially-designed lightweight components.
\item{\bf (III)} We conduct extensive experiments to validate the superiority of our WFD method in comparison to the start-of-the-art models. In particular, given the most challenging traffic data with {\em 17}-tab traces, WFD achieves an mAP of 58.54\%, vs. 5.73\% by the best alternative CDSB \cite{Gong2020}.
Interestingly, against the latest defense GLUE \cite{Gong2020}, WFD even further improves mAP to 80.56\%.
This suggests the proposed \MT~is already more challenging than GLUE-defensed traffic data due to its 
% sequential
{\color{black}consecutive}
single-website visiting assumption. 
In terms of attack speed, WFD achieves 1208.30 Megabytes per second (MBps) with an ordinary NVIDIA 3090 GPU, allowing to attack thousands of Tor users (8516-10113) simultaneously at real time (assuming 16$\sim$19 seconds to load a 2.27 MB sized web-page on average).
% average loading time and 2.27 MB traffic data per web page. 
% the start-of-the-art CDSB series in any conditions even trained with only a little data, with the mAP of 58.54\%(even facing the changing base rates)-vs-5.73\%(their best) dealing with the most difficult $\ell=17$-traces. 
% \item{\bf (V)}
% Against the latest defense GLUE, we achieve better performance than the state-of-the-art CDSB series \cite{Gong2020}, with the highest mAP margin of 72.24\% against the strongest defense situation ($\ell$=17).
% \item{\bf (VI)}
% More importantly, due to its high attack speed of 1208.30 MBps with an ordinary NVIDIA 3090 GPU, our WFD model has the potential to real-time attack thousands of Tor users (8516-10113) simultaneously while they have an average loading time:16-19 seconds for 2.27 MB average traffic data per web page. 
% Hence, our method heralds a new era of the large-scale real-time \MT~even against the secure Tor.
\end{description}

% We organize the rest of the paper as follows. 
% In Section \uppercase\expandafter{\romannumeral2}, 
% we summarize existing multi-tab attacks and compare them with our method. 
% In Section 
% \uppercase\expandafter{\romannumeral3}, 
% we give some preliminaries. 
% In Section \uppercase\expandafter{\romannumeral4}, we present the reformulated WF detection problem. 
% In Section \uppercase\expandafter{\romannumeral5}, we describe our proposed WF detection framework and the corresponding model details. 
% In Section \uppercase\expandafter{\romannumeral6}, we describe the new meaningful performance metrics.
% In Section \uppercase\expandafter{\romannumeral7}, we give the evaluation results under different conditions. 
% The paper is concluded in Section \uppercase\expandafter{\romannumeral8}.
\section{Related Works}
% This section has two parts.
% Firstly, we introduce several representative single-tab WF attacks on Tor.
% Secondly, we discuss all four multi-tab WF attacks and the latest zero-delay defense GLUE. 
\subsection{Single-Tab WF Attack} 
With the anonymized network traffic data of the single web-page as the input, different WF attacks have been proposed.
They can be mainly divided into three categories: traditional \ST ~\cite{Hayes2016,Panchenko2016,Panchenko2011,Wang2014, Wang2013, Wang2016}, deep learning based \ST~ \cite{Abe2016,Rimmer2018,Sirinam2018,Bhat2019} and few-shot \ST~\cite{Sirinam2019, CHEN2021cn, CHEN2021scn}.

{\bf Traditional single-tab WF attacks.} These attack methods mainly use hand-designed feature engineering and %then feed the extracted features into  
traditional machine learning classifiers. 
Representative methods include \emph{k}-NN \cite{Wang2014}, CUMUL \cite{Panchenko2016}, \emph{k}-FP \cite{Hayes2016}, and so forth. 
In the \emph{closed-world} scenario, they can all achieve over 90\% accuracy. 
They have been shown to be effective even in the more difficult \emph{open-world} scenario.
% For the \emph{open-world} scenario, their performance is also excellent.
For instance, \emph{k}-FP can determine 30 monitored web pages from 100,000 unmonitored ones with 85\% true positive rate (TPR), and 0.02\% false positive rate (FPR).
Their success relies on good feature designing by rich domain knowledge, limited to specific characteristics of the training data alone.
Once there is big changes in data pattern, these methods will degrade drastically. 

{\bf Deep learning based \ST.}
Unlike traditional methods,
deep learning models can learn feature representation
from the training data, without the need for manual design, hence more scalable and more friendly.
Importantly, strong results have been obtained.
For example, the DF model \cite{Sirinam2018} achieves 98.3\% accuracy in the \emph{closed-world} setting, 99\% precision with 94\% recall in the \emph{open-world} setting, and even 90.7\% accuracy against the well-known WTF-PAD defense \cite{Juarez2016}.
% Needless to feature engineer, deep learning based WF attacks automate the feature extraction, and implement attack in an end-to-end way.
% In them, DF is noticeable.
% With a sophisticated CNN-based model design, Sirinam et al. \cite{Sirinam2018} proposed the DF attack and achieved 98.3\% accuracy in the \emph{closed-world}, 99\% precision with 94\% recall in the \emph{open-world} and even 90.7\% accuracy against the well-known WTF-PAD defense \cite{Juarez2016}.

{\bf Few-shot \ST.}
Deep models often need a large number of training data, 
which could be difficult to collect in many cases. 
To overcome this challenge, 
% To deal with the data hungry of these deep WF attacks, 
few-shot learning has been introduced 
with representative works including
% WF attacks are studied and achieved excellent performance, such as 
TF \cite{Sirinam2019}, TLFA \cite{CHEN2021cn}, and HDA \cite{CHEN2021scn}.
In particular, TLFA can achieve over 90\% accuracy with only one sample per new website through knowledge reuse of seen websites.

% with knowledge transfer, the TLFA can achieve a performance of over 90\% accuracy, even only trained with only one sample per new website. 
Commonly, all the above works assume well trimmed traces each 
% with a single monitored 
{\color{black}involving a single}
website. Further, they assume singe-tab visiting which is an artificial and unrealistic assumption. As a consequence, these methods are supposed to 
be unsuitable for real-world applications,
as a user often uses multiple tabs in internet surfing. 

% Unfortunately, the above successful WF attacks cannot effectively deanonymize pages if the user surfs in a multi-tab way. 

\subsection{Multi-Tab WF Attack}
In the literature, there are a limited number of works \cite{Gu2015, Wang2016, Xu2018, Gong2020} that study the \MT~problem at varying degrees.
% , although limited in different aspects.

In 2015, Gu et al. \cite{Gu2015} presented the first two-tab WF attack on SSH.
By selecting exclusive features of SSH, such as TCP connections, total per-direction bandwidth and inter-packet time, they exploited a Mahalanobis distance metric to determine whether a  trace is partly overlapped or not.
They further selected fine-grained features to deanonymize the first page and coarse features to identify the second page. 
With a two seconds delay of two pages visits, their attack can classify the first page at 75.9\% TPR and the second one at 40.5\% TPR among 50 websites in a \emph{closed-world} setting. 
Unlike their work, we skip the split decision phase and only focus on detecting monitored trace segments.

In 2016, Wang et al. \cite{Wang2016} firstly studied the two-tab WF attack on Tor.
They noticed the importance of time gap and used a time-based \emph{k}-NN (Time-\emph{k}NN) model to deal with split decision and split finding.
% After splitting decision, they did the split finding of the two-tab traces with their \emph{k}-NN attack \cite{Wang2014}.
% Although their Time-\emph{k}NN can almost successfully do split decision, but its split-finding performance drops dramatically while the two visits are consecutive  or even overlapped.
However, this model struggles when the two visits are consecutive or overlapped
-- a 32\% split accuracy for the overlapped cases.
% They get a 32\% split accuracy for the most difficult overlapped two pages. 
Besides, they first discussed the negative impact of not-appropriate splitting on the successive classification.

In 2018, Xu et al. \cite{Xu2018} advanced the \MT~with the same two-step pipeline but only focused on the first-page attack.
They proposed a BalanceCascade-XGBoost model to find the start of the second page trace
by combining XGBoost \cite{Tianqi2016} and BalanceCascade \cite{BalanceCascade2009}.
Further, they selected the most useful features out from 452 candidate ones with IWSSembeddedNB \cite{Bermejo2014} and modeled the attack with random forests.
% The results showed they 
This model can identify the first page at a TPR of 64.94\% on Tor.
However, it cannot detect the overlapped portion of the first trace.
% However, even for the first trace, they still cannot detect its overlapped part.

In 2020, Gong et al. \cite{Gong2020} studied the \MT~problem from a defense perspective, subject to an assumption of single-tab visiting at a time. They proposed a GLUE defense to glue these non-consecutive traces as consecutive.
The motivation is that \MT~is a much harder problem as compared to \ST.
The FRONT defense is also used to ensure the front privacy of glued traces.
% , based on the difficulty of the \MT~problem.
% Additionally, they used their FRONT defense to ensure the front privacy of the glued traces.
% They also consider the \MT~problem as a spear of shield.
They designed a new Coarse-Decided Score-Based method (CDSB) based on uses a random forest classifier and 511 features to do split decision and split points finding.
This follows the same split and classification pipeline as \cite{Xu2018}.
After trace segments are split, existing \ST~classification models can be then applied,
% the classification is done with the popular single-tab WF attacks, 
such as \emph{k}-NN \cite{Wang2014}, CUMUL \cite{Panchenko2016}, \emph{k}-FP \cite{Hayes2016} and DF \cite{Sirinam2018}.
In case of ordinary consecutive 16 traces,
CDSB can achieve over 45\% TPR and over 41\% precision.
With GLUE, the performance drops significantly, achieving the defense effect.
In this work, we instead show that this GLUE defense is actually
simpler than the realistic \MT~setting. 
This means that multi-tab WF traces are naturally challenging on their own.

% They showed that the CDSB series cannot achieve meaningful results against GLUE of 16 web pages, although it still achieved above 45\% TPR and above 41\% precision while attacking the ordinary consecutive 16 ones. But they don't test their \MT~under the more complicated situation where adjacent visits overlap.

% {\bf Why did previous WF attacks fail while attacking the multi-tab traces?}

% Usually, the data pattern concealed in different single traces of one particular website is similar but different from other website's traces, this is the foundation of the single page WF attack's success.
% Unlike the single trace, the data pattern concealed in the multi-trace is more complex, which cannot be viewed as the simple addition of single data patterns.
% Even more, the overlapping between visits complicates the data pattern further.
% In this situation, the failure of single-tab WF attack is inevitable.

In terms of model design, all the above methods separate the trace split (i.e., localization) and classification components, which disables their joint and interactive optimization during training.
Importantly, it is typical that trace location is largely inaccurate.
This will further challenge the subsequent classification,
resulting in a localization error propagation problem that 
cannot be solved in the classification step.
To address this fundamental limitation,
we introduce a novel Website Fingerprinting Detection (WFD) model
that is end-to-end trainable in a unified deep learning framework.

% For the two-stage multi-tab WF attack, they first try to split the multi-tab trace, and then attack these splitted segments with the single trace WF attack. 
% obviously, the single trace WF attack's performance on the splitted segments is heavily dependent on the quality of split decision and split finding.
% After all, the data pattern concealed in a mixed trace segment is complex,  hard to discover.
% Unluckily, the split is always tough  due to its dependence on distinction or classification.
% The intrinsic logical dependence and contradiction of split and classification lead to its extrinsic failure.

% Overall, although these two-stage solutions have developed the \MT, their performance would  drop sharply while the multi-tab visits become complex and long. 
% We effectively solve the \MT~by unifying positioning and deanonymizing of the monitored trace segments into an end-to-end framework.
\begin{figure}%[h!]
    \centering
    \includegraphics[scale=1]{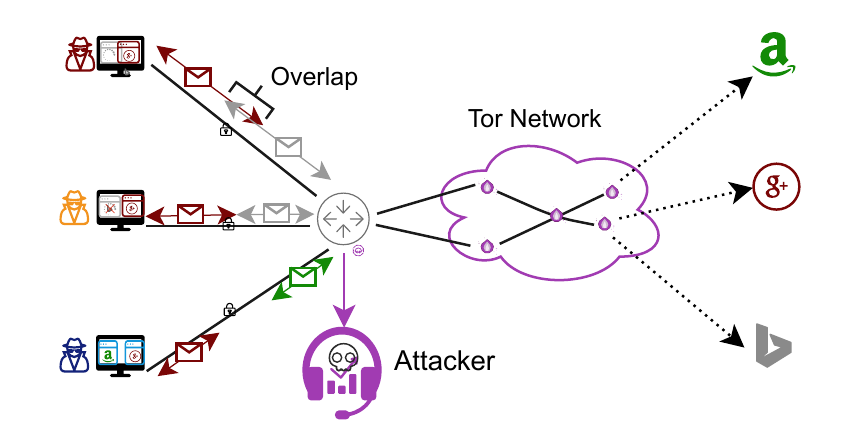}
    \caption{Illustration of data flow between a user and 
several websites with a Tor network in-between, producing single-tab or multi-tab traces traces.
Especially, we consider an attacker taps on the key router and carries out a large-scale real-time \WFA, including \ST~and \MT.
% even considering the difficult anonymous visits with multi-tab.
Suppose Google is one  monitored website, we aim to detect all traces with Google website. 
}
\label{fig:threat_model}
\end{figure}

\section{Preliminaries}
\subsection{Positioning a Threat Model} 
Aiming at the user's browsing privacy, \WFA~is a passive, undetectable and unpreventable traffic analysis.
It does not modify, drop, or delay any packets in the traffic. 
Different from previous \WFA~works limited to the local environment such as LAN, we consider that a WF threat model can settle on a core router for large-scale real-time attacking, as illustrated in Figure \ref{fig:threat_model}.

% should not be limited to the local environment such as LAN, but can be settled on the core router to deploy a large-scale real-time attack, shown in Figure \ref{fig:threat_model}.

\subsection{Common \WFA~Settings} 
% To simplify the realistic \WFA~problem, 
Existing \WFA~works mainly consider these settings:
% evaluate their performance under mainly three assumptions:
\emph{single-tab} vs \emph{multi-tab},  \emph{closed-world} vs \emph{open-world}, and \emph{without-} vs \emph{with-} background traffic. 
We describe their key aspects below.

The majority of existing \WFA~works consider the single-tab situation,
that is, each time a user visits only a single website with a single tab in the browser. This is clearly not common user behaviors.
Instead, multi-tab situations would be the norm in real-world case.
Hence, \MT~is a more realistic problem setting. 

The \emph{closed-world} setting assumes 
that only a fixed number of websites are visited, i.e., the set of {\em monitored websites}.
In real world, this is often not true.
% are available, hence we do not consider it in this paper.
The \emph{open-world} setting eliminates this restriction
by considering the presence of {\em unmonitored websites} in the traffic data.
% The base rate is defined as {\color{blue}the ratio} between the visits of unmonitored and monitored websites.
{\color{black}An important notion is the base rate defined as the rate at which the user visits a page from the set of monitored websites.}
It is user specific and relates the evaluation in the \emph{open-world} case.
Setting a too low base rate will lead to overly optimistic results,
i.e., the ``base rate fallacy'' phenomenon \cite{Juarez2014}.
% is consistent with the realistic setting, in which the user's visit is not restricted to the {\bf monitored websites set} but expanded to almost unlimited unmonitored websites.

% For the number of webpage visits, we concern the more realistic \MT~problem. 
% Each time one visit to only a single website is obviously not frequent, and the multiple visits to different websites spontaneously are the default state of user behavior.

% For the websites visiting, the \emph{closed-world} assumes only a limited number of websites are available, hence we do not consider it in this paper.
% The \emph{open-world} is consistent with the realistic setting, in which the user's visit is not restricted to the {\bf monitored websites set} but expanded to almost unlimited unmonitored websites.

\begin{myDef}
\label{Monitored Websites Set}
{\bf A set of monitored websites.} They are the target websites $\mathcal{S}$ an attacker wants to monitor.
Each website $\bm{w}$ in $\mathcal{S}$ can be labeled as a specific class.
So \ST~is often formulated as a classification problem.
% With a fixed order, the \emph{i}-th website $\bm{y}$ in $\mathcal{S}$ can be labeled as $i$. 
{\color{black}For all the unmonitored websites in the open-world scenario, 
they are usually collectively labeled as a single class
different from all the classes assigned to monitored websites.
% their labeling is no sense.?/For the open-world scenario, this labeling applies to these unmonitored websites but only with the same class $n+1$  if the size of $\mathcal{S}$ is $n$.?
}

% In the open-world scenario,  each of the websites $\bm{y}$ out of the monitored set $\mathcal{S}$ are samely labeled as $n+1$ if the size of $\mathcal{S}$ is $n$.
\end{myDef}

% The \emph{open-world} evaluation results are often criticized to be overly optimistic due to the too low base rate between visits of unmonitored and monitored, which actually depends on the user.
% To avoid the above "base rate fallacy" \cite{Juarez2014}, this ratio should not be settled to too low in experiment setting.

The background traffic is generated often due to file downloading, online music streaming, and so on. Given that the current Tor browsing is relatively slow \cite{WangBlast2020},
these background activities are usually not considered as they will further slower the flow of traffic data and reduce the user experience.

% we do not consider this issue because the current Tor browsing is relatively slow \cite{WangBlast2020}, and would be slower if browsing with Tor background traffic.

\section{Real-World Multi-Tab WF Attack}
% \subsection{Problem Reformulation}

\subsection{Internet User Behaviors}
Adjacent website visiting is one of the most challenging situations for \WFA.
There are three typical types of user behaviors (Fig. \ref{fig:threat_model}) as below.
% As shown in Fig. \ref{fig:threat_model}, 
(1) The blue-colored user dwells on an old page for a while before opening a new page, yielding a traffic sequence with {\em two separated and sequential single-tab traces}.
(2) The yellow-colored user reloads a new page in the current tab that stops the loading of the old incomplete page, yielding a traffic sequence with {\em two consecutive} without a noticeable  time gap.
This is a {\em multi-tab trace} with individual traces all clean.
(3)
The red-colored user opens a new page in a new tab while the old page is in loading, yielding  a traffic sequence with {\em two overlapping traces}.
This is a more challenging {\em multi-tab trace}, which has never been
systematically investigated in the literature.
To facilitate understanding,
we define several key concepts in follows.
% the overlapped part 
% being the 
%in which there are both {\bf clean trace segments and mixed ones}.

\begin{myDef}
\label{def:single_trace}
{\bf A single-tab trace.} A single-tab trace is a sequence of packets produced 
in the loading process of a web browser when a user pays a visit to one website.
It is also known as {\em a single trace}.
% by one visit of a user to one website, corresponding to the webpage loading process in the browser.
\end{myDef}

\begin{myDef}
\label{CMT}
{\bf A multi-tab trace.}
A multi-tab trace comprises more than one traces with or without overlapping,
produced in the loading process of a web browser when a user opens multiple tabs to 
visit multiple websites concurrently.
% of consecutive or overlapping visits to websites as if a long single trace, corresponding to the loading process of webpages in  browser's multiple tabs. 
\end{myDef}

\begin{myDef}
\label{cdts}
{\bf Trace segment.}
A trace segment is a consecutive sub sequence of one (i.e., clean) 
and multiple (i.e., mixed) traces.
% If it is all from one single-tab trace
% A clean trace segment is a trace segment whose packets are all from one single-tab trace, and "mixed" means its some packets are from others because its some part overlaps with others.
\end{myDef}
% \subsection{Definitions}
% We give the definitions of monitored set, cell, burst, single trace, multi-tab trace, trace segment, ground-truth, cell segment, anchor segment, proposal segment and intersection over union of segments:

% \begin{myDef}
% \label{Monitored Websites Set}
% {\bf Monitored set.} A monitored set is the websites $\mathcal{S}$ the attacker wants to monitor. 
% With a fixed order, the \emph{i}-th website $\bm{y}$ in $\mathcal{S}$ can be labeled as $i$. 
% In the open-world scenario,  each of the websites $\bm{y}$ out of the monitored set $\mathcal{S}$ are samely labeled as $n+1$ if the size of $\mathcal{S}$ is $n$.
% \end{myDef}

% \begin{myDef}
% \label{cs}
% {\bf Cell segment.}
% A cell segment is a trace segment whose length is fixed as the down-sampling rate of the feature extractor, corresponding to one feature vector in the extracted sequence of feature vectors.
% \end{myDef}
% \begin{myDef}
% \label{A-TS}
% {\bf Anchor segment.}
% An anchor segment is a trace segment whose length is pre-fixed for the suggestion of high-quality proposal segments. 
% \end{myDef}

\subsection{Problem Definition}
Given a raw traffic data (i.e., a long untrimmed multi-tab trace), the {\bf \em objective} of \MT~is to identify any single-tab traces of all monitored website,
including their start, end, and website.

% develop a WF attack model which can answer:
%  which and where? -- For each individual full monitored trace ({\bf ground-truth}), which monitored website the user is visiting, and where is it in the raw traffic data of multi-tab traces? 
% The objective of which and where means taking full monitored traces (ground-truth) as a specific object to detect.
% Due to the complexity and uncertainty of data patterns in mixed trace segments, we select the clean monitored trace segment as the evidence.
% This is essentially viewing the \ST~problem from the perspective of detection.

\begin{myDef}
\label{GT-TS}
{\bf Ground-truth.}
In \MT, a ground-truth is defined as a full monitored trace with
the index of the first $s$ and last $e$ packet, and a website label 
${y
% \not=n+1
} \in S$, in a raw multi-tab trace.
Its center point is $c = (s + e) / 2 $, and the length is $l = e -s$.

% , and $y$ is its website label in the monitored set $S$.

% and whose first and last packets are all from the website $y$.
% Its position and label can be denoted as $(p,y)$ where $p = (s,e)/(c,l)$ is its position representation, $s/e$ is the index of the first/last packet, $c = (s + e) / 2 $ is the index of its center point, $l = e -s $ is its length, and $y$ is its website label in the monitored set $S$.
% A clean ground-truth is the longest clean trace segment of the ground-truth.

\end{myDef}

In order to train a \MT~model for detecting all the ground-truth traces from a given multi-tab trace,
we often collect a labelled training set $D = \{{x}_i)\}_{i=1}^N$ with each trace ${x}_i$ associated with a set of $J$ ground-truth traces 
{\color{black}
% $Y_i = \{y_i\} = \{(s_i, e_i, w_i)\}$
$Y_i = \{y_{ij}\}_{j=1}^J = \{(s_{ij}, e_{ij}, w_{ij})\}_{j=1}^J$}.
%
% to get their positions $p$ and website labels $y$.
% Now, we give the formal definition of \MT~problem: 
% how to build a model 
% % $\epsilon$ 
% to detect all the ground-truths in the multi-tab trace to get their positions $p$ and website labels $y$.
% For model training, a labelled training set $D = \{{x}_i)\}_{i=1}^N$ is often provided where ${x}_i$ represents the raw traffic data of \emph{i}-th sample (multi-tab trace)  in which 
% % the position $p$ and label $y$ of 
% all the monitored trace segments
% are given as the ground-truths.
In attack (i.e., model inference), given a test multi-tab trace $\tilde{x}_i$ a trained model
would output a set of $K$ top-scoring {\em candidate traces} {\color{black}$\hat{Y}_i = \{\hat{y}_{ik}\}_{k=1}^K$}.
Ideally, {\color{black}$\hat{Y}_i$} matches the ground-truth traces {\color{black}$\tilde{Y}_i$} exactly.
This however is rarely possible.
Typically, we hope that {\color{black}$\hat{Y}_i$} has high overlap (Eq. \eqref{eq:map_lamda})
with {\color{black}$\tilde{Y}_j$}.

% For model test/attack, given a multi-tab trace as input, if there exist some ground-truths,
% the trained model would output the {\bf proposal segments} as the possible predictions.
% For their positions, the {\bf IoUT} values between the proposal segments and their ground-truths should be optimized as high as possible.
% For their website label probabilities, these ones on the index of ground-truths' website label should be optimized as high as possible.
% whose positions and website labels would be predicted .

\begin{myDef}
\label{P-TS}
{\bf A candidate trace.}
A candidate trace is a prediction 
% $\hat{y}_j=(\hat{c}_j, \hat{l}_j, \hat{w}_j, \hat{s}_j)$
{\color{black}$\hat{y}=(\hat{c}, \hat{l}, \hat{w}, \hat{s})$}
outputted by a \WFD~model, 
including a specific website label {\color{black}$\hat{w}$},
the center point position $\hat{c}$, the length $\hat{l}$,
and a detection score $\hat{s}$.

% its maximum probability Pr$_{y}$ of the specific website label ${y}$ , its center point position ${c}$ and its length ${l}$.
%
% trace segment the WFD model proposes as a possible ground-truth,
% % based on the anchor segment, 
% aiming to predict its maximum probability Pr$_{y}$ of the specific website label ${y}$ , its center point position ${c}$ and its length ${l}$.
\end{myDef}

\begin{myDef}
\label{IoUT}
{\bf Intersection over union of Traces (IoUT).}
The intersection over union of two traces is defined as the ratio between the length of their common segments to the length of their union.
% the shortest segment containing the above two traces.
\end{myDef}

\begin{figure}[h]
    \centering
    \includegraphics[scale=1]{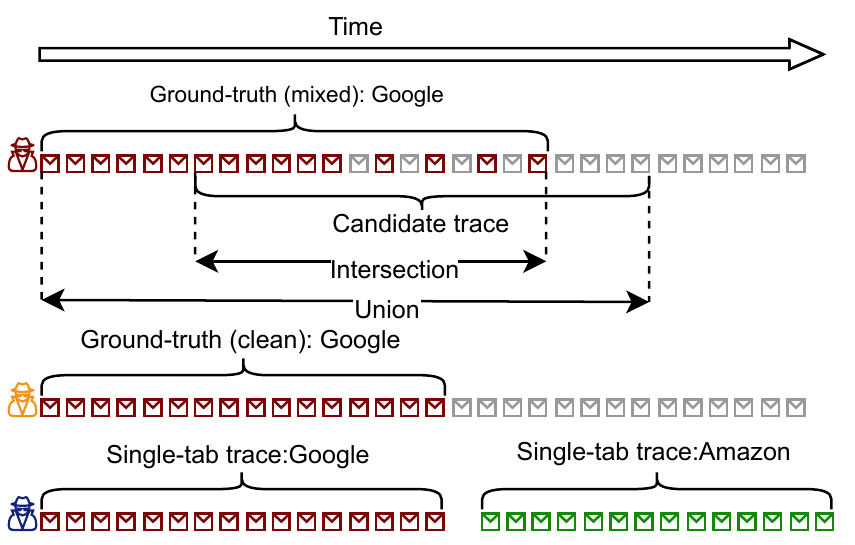}
    \caption{Illustration of full traces and trace segments: single-tab trace, multi-tab trace (consecutive or overlapped), ground-truth (clean or mixed), candidate trace, and the Intersection and Union of IoUT. In this example, we consider Google (red) as one monitored website, and Amazon (green) and Bing (grey) as unmonitored ones.
    }
\label{fig:seg}
\end{figure}

To better understand the above concepts,
we provide an example in Figure \ref{fig:seg}.

% As shown in Figure \ref{fig:seg}, for the red/yellow-colored user, his/her visits produce the overlapped/consecutive multi-tab trace, in which the ground-truth (red, mixed/clean) are labeled.
% For the ground-truth, we use the indexes ${s,e}$  of its first packet and last packet to calculate its center point ${c}$ and length ${l}$.
% While training or attacking, the \MT~model would output some candidate traces to predict these ground-truths.
% The IoUT would be used to evaluate the positioning accuracy.

% Taking such a detection perspective, the \MT~problem can be viewed as a two-task optimization problem: one task is location/regression, and another is classification. We would set the corresponding attack goals and perfromance metrics for the above two tasks.
% We would optimize these two tasks in a unified and mutually beneficial way.

\begin{figure*}[!h]
    \centering
        \includegraphics[scale=0.99]{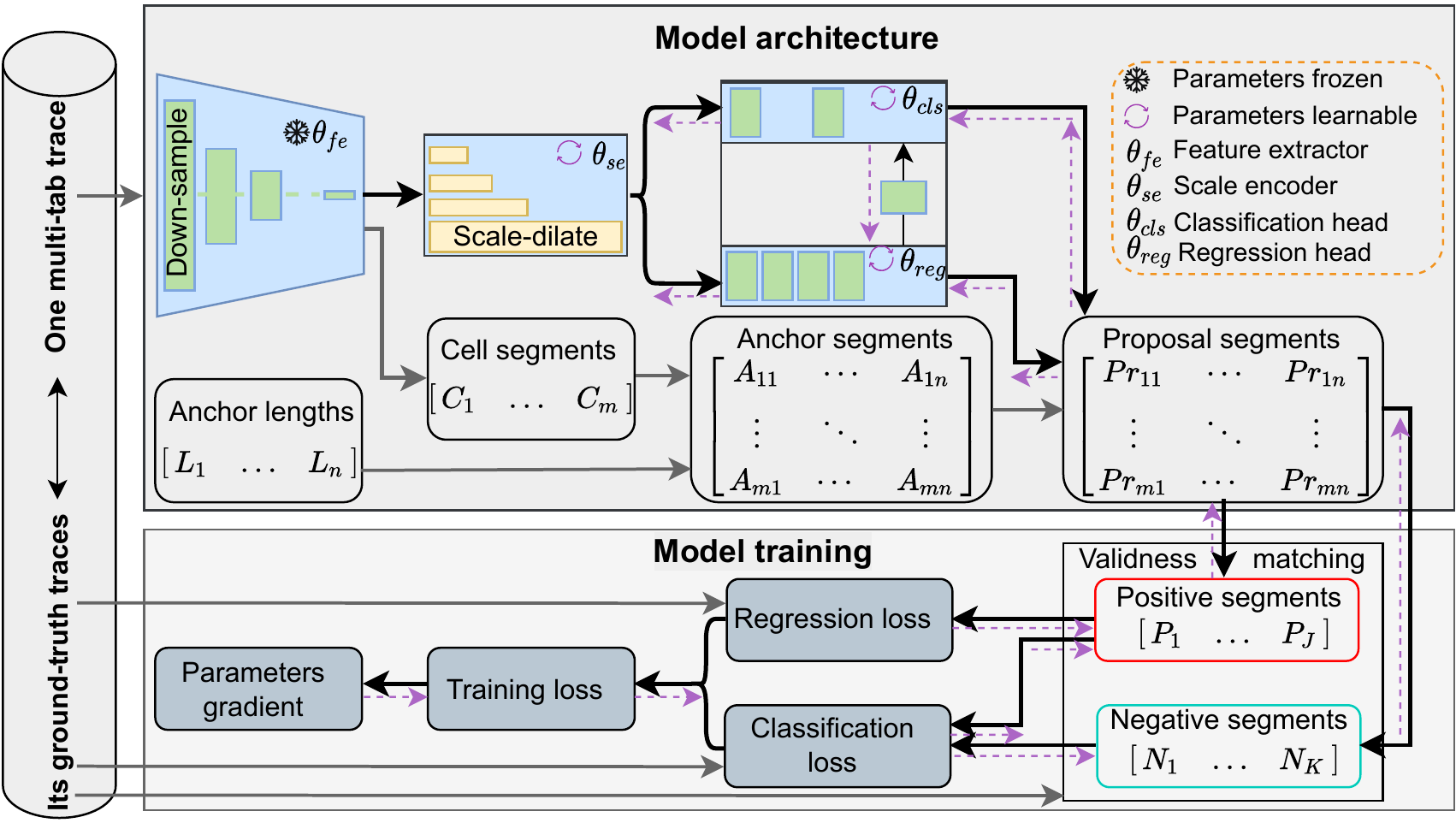}
    \caption{
    Schematic overview of the proposed {\bf \em Website Fingerprinting Detection} (WFD) model.
    WFD solves the \MT~problem by jointly optimizing the classification task and location task end to end.
    Firstly, the feature extractor $\theta_{fe}$ down-samples an input multi-tab trace to a $m$-length feature vector sequence.
    Each feature vector corresponds to a cell segment.
    For each cell segment, $n$ anchor segments are placed.
    Then, the scale encoder $\theta_{se}$ learns rich information at different scales. 
    With the position offsets and website probabilities predicted by the regression head $\theta_{reg}$ and the classification head $\theta_{cls}$, $m\times n$ anchor segments are converted into $m \times n$ proposal segments.
    % These proposed segments would be used for both training and attacking processes.
    In model training, we match these proposed segments with the ground-truths, and obtain $J$ positive segments and $K$ negative segments.
    All these segments are used to calculate the classification loss.
    However, only the positive segments are used for the calculation of regression loss.
    The summed loss is used for model parameter update via SGD.
    }
    \label{fig:wfd_model}
\end{figure*}
\subsection{Performance Metrics}
In this section, we introduce a new set of performance metrics for evaluating both the {\em accuracy} and {\em efficiency} of a \MT~method.
% This is because, the precision and recall metrics designed for conventional \ST~cannot fully reflect the requirements of \MT.
% For accuracy, we suggest the mean Average Precision (mAP) with varying IoUT threshold values to accommodate a collection of different demanding degrees.
% For efficiency, we suggest two speed metrics: the training time and the attack speed
% in Megabytes per second (MBps).
% We define their formulations next.

\subsubsection{Accuracy metrics}
% In this section, we firstly define precision and recall with threshold value, then average precision (AP), and lastly the mAP.
The precision and recall metrics designed for conventional \ST~cannot fully reflect the requirements of \MT.
{\bf Precision and recall at thresholds. }
Given a predicted trace and a ground-truth,
if their IoUT value exceeds a predefined threshold $\lambda$ (e.g., 0.5), and the maximal classification probability corresponding to the ground-truth website $i$ and exceeds a predefined threshold $\tau$, this prediction is then viewed as a true positive ($TP^{i}_{\lambda}\left(\tau\right)$).
Otherwise it is a false positive ($FP^{i}_{\lambda}\left(\tau\right)$).
If there is no predicted trace for a ground-truth, it is viewed as a false negative ($FN^{i}_{\lambda}$).
The precision and recall metrics for website $i$ are then defined as:
\begin{equation}
    precision^{i}_{\lambda}\left(\tau\right) = \frac{TP^{i}_{\lambda}\left(\tau\right)}{TP^{i}_{\lambda}\left(\tau\right) + FP^{i}_{\lambda}\left(\tau\right)}
\end{equation}
and
\begin{equation}
    recall^{i}_{\lambda}\left(\tau\right) = 
    \frac{TP^{i}_{\lambda}\left(\tau\right)}{TP^{i}_{\lambda}\left(\tau\right) + FN^{i}_{\lambda}}
\end{equation}

% For a proposed segment the monitored website label of whose ground-truth is $i$, if the IoUT value  exceeds a predefined one $\lambda$ (i.e., 0.5), and the \emph{i}-th classification probability value exceeds a pre-defined value $\tau$, it can be viewed as a true positive ($TP^{i}_{\lambda}\left(\tau\right)$), otherwise a false positive($FP^{i}_{\lambda}\left(\tau\right)$).
% While there is no proposed trace segment for one ground-truth, it is viewed as a false negative ($FN^{i}_{\lambda}\left(\tau\right)$).
% The true negative (TN) is not used.
% With the above ($TP^{i}_{\lambda}\left(\tau\right)$, $FP^{i}_{\lambda}\left(\tau\right)$ and $FN^{i}_{\lambda}\left(\tau\right)$, the corresponding $precision^{i}_{\lambda}\left(\tau\right)$ and $recall^{i}_{\lambda}\left(\tau\right)$ can be easily defined.

\noindent{\bf Average precision.}
By varying the classification threshold $\tau$,
we can draw a series of precision-recall score pairs.
To summarize their performance, we consider the area under the precision-recall curve as a metric, namely Average Precision (AP).
% The precision-recall curve can show the trend of precision with the change of recall according to different $\tau$. 
% For simplicity, the area under the precision-recall curve is proposed, named average precision (AP).
Formally, the AP for website $i$ at a IoUT threshold value $\lambda$
is defined as:
% , the AP for the website $i$ is defined as the weighted sum of precisions at each $\tau$ where the weight is the increase of recall as follows:
\begin{equation}
\begin{aligned}
AP^{i}_{\lambda} &= 
\int_{\tau=0}^{1} Precision^{i}_{\lambda}\left(\tau\right)\, d(Recall^{i}_{\lambda}\left(\tau\right))
\end{aligned}
\end{equation}

\noindent{\bf mAP.}
To summarize the AP scores over a set of monitored websites $S$,
we define the mean average precision (mAP$_{\lambda}$) metric as:
% To represent the 
% average effectiveness of the \MT~on the specified website,  the mean average precision (mAP) is proposed,
\begin{equation}
\label{eq:map_lamda}
\begin{aligned}
mAP_{\lambda} &= 
\frac{1}{|S|}
\sum_{i=1}^{|S|} Precision^{i}_{\lambda}\left(\tau\right)
\end{aligned}
\end{equation}
where $|S|$ is the number of monitored websites.
For a specific IoUT threshold, e.g., $\lambda=0.50$,
its corresponding mAP is denoted as mAP$_{.50}$.
By default, we suggest a range of IoUT threshold
to $[0.5, 0.9]$ with a step 0.05.
Finally, we average all the mAP$_{\lambda}$ 
to obtain a single mAP score for overall performance.
% In experiments, with different IoUT threshold values such as 0.5, 0.75, mAP$_{50}$, mAP$_{75}$ are given.
% While the IoUT threshold values are set up from 0.5 to 0.9 with a step 0.05, we get the mean of different mAP$_{\lambda}$ values, noted as mAP.

\subsubsection{Efficiency metrics}
For {\color{red}the evaluation of} efficiency, we suggest two speed metrics: the training time and the attack speed
in Megabytes per second (MBps).
The training time can be easily measured on a fixed machine.

% we detail the metric of attack speed: Megabytes per second.

\noindent{\bf Megabytes per second (MBps). }
Although the number of traces per second can also be used to evaluate the attack speed, this does not reflect the trace length.
%
% it is needed to specify the number and length of single-tab traces contained in the multi-tab trace.
To overcome this issue, we suggest using Megabytes per Second (MBps) 
as the attack speed metric.
% to evaluate the WF attack speed in an intuitive way.
In \MT, MBps measures the amount of the anonymous network flow in the unit of in Megabytes a \WFA~method can detect per second.
%
% In such a way, the attack speed can be compared fairly.

\subsection{Remarks}
\label{sec:remarks}
% Unlike 
% single-tab WF attack,
% the multi-tab WF attack has not been well studied.
% In this section, we focus on the specialized goals of \MT: the core classification and forensics, bootstrap time, and attack speed.

{\bf Classification vs. forensics.} 
\ST~is essentially a classification problem
of clean, manually trimmed single-tab traces.
This is a simplified form of \MT.
In forensics, one needs to process untrimmed multi-tab traces instead
and detect the location of singe-tab traces with monitored websites,
without any manual trimming.
% For the single-tab trace, the whole trace can be used as evidence to support the attack/classification.
% However, for the \MT, only announcing the final classification results (which monitored websites the user visits) is not enough.
% The attacker needs to know where are the ground-truths.
% Through the forensics, the predicted position can be viewed as evidence of a \MT~success.
% In short, we need to know which and where, corresponding to classification and forensics respectively.

\noindent{\bf Bootstrap time and attack speed. }
The \WFA~time is critical for an attacker, including bootstrap time \cite{Sirinam2019} and attack time.
The former is the time required to build a ready-to-use \WFA~model, including data collecting time and \WFA~model training time.
% After bootstrap, an attacker can get a ready-to-use \WFA~model.
% obviously, a shorter bootstrap time means a better \WFA.
The attack speed is also important due to the potentially
infinite amount of WF data.
% As everyone knows, information has timeliness.
% Similarly, privacy also has timeliness.
% A high-speed WF attack can compromise a user's privacy at real-time, even many user's on a large scale.
%
In general, we aim to develop fast-to-train, fast-to-attack, and more accurate \WFA~models.

% Overall, with precondition of accurate classification and forensics, a quick-prepared and quick-attacking \WFA~model would be more dangerous.

\section{End-to-End WF Detection}
% Since the multi-tab WF problem has been reformulated as a detection problem, an appropriate design of the WF detection framework for WF detection is necessary and important.
In this section, we present our WFD model, including its motivation, overview,  
model architecture, model training,  model testing (attack) and model optimization.
% segment proposal, train process, test/attack process, and optimization.

% \subsection{ Inspiration of WF Detection Framework}
\subsection{Motivation}
% We take inspiration from object detection in computer vision.
% In the realistic multi-tab open-world scenario, the attacker's focus is also the visits to monitored websites, samely as the single page visit.
% To be more concrete, the goal of a multi-tab WF attack is to do the classification and forensics of monitored trace segments concealed in the long multi-tab trace. 

We associate the \MT~problem with object detection in computer vision due to their remarkable success achieved \cite{He2014,Girshick2015,ren2015faster, Lin2017,Redmon2016,SSD2016,Focalloss2018, chen2021look}.
This is based on our {\em novel analogy} that
that each individual single-tab trace of any monitored website can be viewed as an object of interest in an image, whilst all the other traces 
as the background of the image.
This perspective opens a door to explore the rich experiences
of object detection models.
However, it is still non-trivial to develop a strong \MT~model
in analogy of object detectors due to different problem specific challenges.

Considering the demand of accuracy and efficiency, we choose the one-stage solution \cite{chen2021look} as the architecture of our WF detection, which can solve the regression problem and the classification problem in a unified way, with a high speed.

\subsection{Overview}
For \MT, we design a novel {\bf \em Website Fingerprinting Detection} (WFD) model, as illustrated in Figure \ref{fig:wfd_model}.
It is end-to-end learnable, taking as input a raw multi-tab trace,
and outputting a set of candidate traces.
Both trace location and classification are jointly
optimized with strong synergy in-between realized.

% supervised learning that simulates the exactly the same scenario of \MT~with the help of an end-to-end architecture.
% By unifying the optimization of location and classification, our WFD model well addresses the two sub problems of \MT.

% In such a way, the WFD model can be more effectively trained with a reasonable difficulty.
% In this section, we give out an overview of our WFD model, including the
% model architecture, model training, and model testing (attack).

% {\bf Segment proposal. }
{\bf Model Architecture. }
The proposed WFD model consists of three functional components:
feature extraction, 
scale perception, and 
localization and classification.
%
% the unified prediction of location offsets and website probabilities.
% for the segment proposal: the first focusing on the feature extraction, the second focusing on the scale perception, and the last focusing on the unified prediction of location offsets and website probabilities.
%
Specifically, 
% The procedure of its segment proposal can be described as follows:
% With the lengths of the ground-truths as input, the lengths of anchors are clustered.
given a raw traffic datum of a whole multi-tab trace, the feature extractor generates a feature vector sequence and cell segments.
Anchor segments are then generated based on cell segments and anchor lengths.
% With the feature vector sequence as input, t
The scale encoder further perceives different scales to learn scale-sensitive information for feature representation.
% outputs a new scale feature vector sequence in which different scale-sensitive features are contained.
At the end, a regression head outputs the offsets of proposal segments and a classification head conducts website classification,
i.e., the two-head predictor.
These outputs are used to generate candidate traces.

% With anchor segments and the offsets of proposal segments, the proposal segments are generated.

{\bf Model training.}
The main steps of model training can be summarized as follows:
(1) Matching the proposal segments with the ground-truth;
(2) Obtaining valid positive and negative segments;
(3) For positive segments, computing the regression loss of the predicted offers against the ground-truth location;
(4) For all valid proposal segments, computing the classification loss of the predicted probabilities against the ground-truth label;
(5) Summing up the above two losses, computing the gradient of all learnable parameters for parameter update.
% Matching with the ground-truths, the validness of proposal segments are checked and these valid 
% positive-vs-negative segments are ouputed.
% With positive segments and the ground-truth, the regression loss is calculated.
% With the predicted probabilities of valid proposal segments and the ground-truth label, the classification loss is calculated.
% With the above two losses, the gradient of these learnable parameters is  calculated and used for the parameter update.
% Notably, we pre-train the first component beforehand with a large-scale single trace dataset and  then freeze its parameters so that only a very little fraction of parameters from the other two components needs to be updated.

{\bf Model testing (attack).}
Given a raw traffic datum of a multi-trace, the trained model can output a number of candidate traces.
% These proposal segments are too many.
As not all of them are good, we select the those with top classification scores as the output.
% With them, we filter out these invalid ones and use the left ones as the final candidate traces.

% All the above components and data flows are illustrated in Figure \ref{fig:wfd_model}.

% \subsection{Segment proposal}
\subsection{Model Architecture}
% A proposal segment is actually predicted as a candidate for some monitored trace segments in the multi-tab trace.
% For model architecture, we first introduce the main components of our WFD model, and then detail the important procedures of segment proposal.
% The segment proposer is mainly designed to propose possible monitored segments for any multi-tab trace input.
\subsubsection{Main components }
% Our WFD model architecture contains three main components: feature extractor, scale encoder and two-head predictor.

{\bf Feature extractor. }
% For solving the \WFA~problem, w
A strong model is needed to extract a powerful feature representation for each raw traffic datum of multi-tab traces.
We choose the 1-dimensional (1D) variant of residual convolution neural networks (ResNets) \cite{He2016}.
% In all types of CNN, we simply choose the ResNet \cite{He2016} series as the feature extractor of our new feature representation.
From a multi-tab trace with length $l^{multi}$, the feature extractor with a temporal down-sampling rate $r_{ds}$ can efficiently extract a feature vector sequence with a length of
$m=\lfloor l^{multi}/r_{ds}\rfloor$.
Each feature vector corresponds to one cell segment.
For the \emph{i}-th cell segment, its center point is $c=(i-1)\times m + m/2$ and its length is $m$.

\noindent {\bf Scale encoder. }
Given a feature vector sequence, we let feature vector take charge of predicting the ground-truths at different lengths.
This allows to learn rich scale information.
%
% To enlarge the receptive field for more contextual information,
One intuitive method is that we deploy a stack of convolutional layers
and extract $n$ feature sequences at multiple layers.
However, this is less efficient and computationally demanding.
To overcome this problem,
we propose using dilation convolutions with different scales \cite{DenOord2016, yu2016multi-scale, Bhat2019}.
% However, one feature vector extracted by the CNN feature extractor only contains feature of its cell segment.
% This problem can be solved by extracting $n$ feature sequences from different layers of CNN.
% obviously, this solution is complex, time-consumed, and resource-consumed.
% With dilation, we can easily perceive segments with different scales.

% However, one feature vector extracted by the CNN feature extractor only contains feature of its cell segment.
% This problem can be solved by extracting $n$ feature sequences from different layers of CNN.
% obviously, this solution is complex, time-consumed, and resource-consumed.
% With dilation, we can easily perceive segments with different scales. 

Specifically, we adopt the 1D residual dilation block as the basic component.
Except the shortcut, such a block consists of three consecutive convolutions: a channel reduction convolution, a dilation convolution for enlarging the receptive length, a channel recovery convolution.
By stacking multiple such blocks, different scales of segments can be perceived.
As a consequence, more scale-sensitive information can be captured
in feature representation learning.
% added into the feature vector sequence.
%
For cost-effective design, the number of dilation blocks
needs to be tuned, along with the dilation rates.
Our optimal design has 4 blocks with dilating rates [2,4,6,8].

% Too few 1D residual dilation blocks are not enough to perceive all these ground-truth lengths, but too many ones are redundant.
% We find that the design of 4 blocks with dilating rates [2,4,6,8] performs best.

{\bf Two-head predictor.} 
One head is responsible for the classification task, and another head for the regression task.
We adopt the anchor design \cite{ren2015faster}.
Specifically, we place $n$ anchor segments at each feature vector (i.e., a cell segment), resulting in a total of $m\times n$ anchor segments.
For all the anchor segments, the classification head outputs the probability vectors in shape of $ m\times n\times|S|$, and the regression head outputs the position offsets in shape of $ m \times n \times2$.
Further, we deploy a convolution layer as a shortcut connection from the regression head to the classification head for facilitating learning. 
% This is because we think a good classification/recognition is beneficial to the regression/location. 
% For the two-head predictor, one head is responsible for the classification task, and another head for the regression task.
% For these $m\times n$ anchor segments, the classification head outputs the probability vector whose size is $ m\times n\times|S|$, and the regression head outputs the position offsets whose size is $ m \times n \times2$.
% Furthermore, we  set a convolution layer as the shortcut connection from the regression head to the classification head because we think a good classification/recognition is beneficial to the regression/location. 

\subsubsection{Key designs}

There are three key designs involved in generating proposal segments,
as described below.
% : anchors-length generating, anchor segments generating and proposal segments generating.

{\bf Anchor lengths. }
We define a set of fixed anchor lengths as the base, from which the predicted offsets are translated to the absolute positions of proposal segments. 
We cluster the lengths of ground-truth traces by the $k$-means,
and obtain $n$ anchor lengths.

% While we set the centers of k-mean as $K$, we would get $K$  anchor-lengths, preparing for the generation from cell segments to their anchor ones.

{\bf Anchor segments. }
For each cell segment, we assign $n$ anchor segments.
For a multi-tab trace with $m$ cell segments, % divided by the feature extractor,
a total of $m\times n$ anchor segments can be resulted.
For each anchor segment, its center point is at the corresponding cell segment.
and its length is one of the $n$ anchor lengths.

{\bf Proposal segments. }
For an anchor segment, we generate one and only one proposal segment.
% would be generated by the proposal generator.
Hence, there are a total of $m\times n$ proposal segments generated.
Each proposal segment shares the classification output of the corresponding anchor segment, and its position is calculated by the predicted position offsets and the anchor segment's position.

% For these proposal segments and their corresponding anchor segments, their predicted website label probability vectors are the same.
% The positions of these proposal segments can be calculated by the predicted position offsets and the positions of anchor segments.
Taking a proposal segment $ps$ as example, suppose the length and center point position of the corresponding anchor segment $as$ are $l_{as}$ and $c_{ps}$, the predicted length offset and center point position are $d_{c}$ and $d_{l}$, then its length and center point can be calculated as follows:
\begin{equation}
\begin{aligned}
c_{ps} &= \sigma(d_{c}) + c_{as} \\
l_{ps} &= l_{as} \times e^{d_{l}}
\end{aligned}
\end{equation}
where $\sigma$ is the sigmoid function.

\subsection{Model Training}
% Since the training process has been introduced above, 
For model training, we give details for validness, objective loss function, and parameter update.

{\bf Validness. }
The proposal segments cannot be used to calculate the objective loss until passing through a validation process.
Firstly, for each proposal segment, we match it against all the ground-truths according to the IoUT metric.
Then, for each ground-truth, we sort its matched proposal segments according to their $\emph{L1-}$distance, and take the $\emph{k}$-nearest neighbors as positive segments, and the remaining as negative segments.
During training, the similarity between a positive segment and its ground-truth in website class probability and location offset is maximized.  
On the contrary, the similarity between a negative segment and its ground-truth
is minimized. 
We discard invalid segments, including positive ones with too low IoUT and negative ones with too high IoUT.
This is because they would harm the model training.

% Now, all the valid positive and negative segments are prepared well for the following losses calculation.

% For a positive segment and its ground-truth, the distance is short, the data pattern learned is similar, its website class probability and location offset can be easily optimized to be similar.
% For a negative segment and its ground-truth, the distance is long, the data pattern learned is different, its website class probability can be easily optimized to be different, and its location offsets are not considered.
% Significantly, we discard the invalid ones -- the IoUT-value too low positive ones and the IoUT-value too high negative ones, because they  would harm the train of WFD model.
% Now, all the valid positive and negative segments are prepared well for the following losses calculation.

{\bf Objective loss function.}
For one ground-truth with website label $w$, the $\emph{w}$-th estimated probability of its positive segments and negative ones is used to calculate the classification loss. 
For the regression loss calculation, only the positive ones are used.
Overall, for a multi-tab trace $x_i$ and its $J$ ground-truths $Y_i = \{y_{ij}\}_{j=1}^J$, the final loss $\mathcal{L}_i$ can be written as follows:
\begin{align}
\mathcal{L}_i &= \sum_{j=1}^{J}\left[\mathcal{L}_{cls}({y_{ij}}) + \mathcal{L}_{reg}({y_{ij}})\right]\\
\mathcal{L}_{cls}({y_{ij})} &=
     \sum_{k=1}^{K_{ij}^{P}}L_{cls}\left(y_{ij}, \hat{y}_{ijk}^P\right) + \sum_{k=1}^{K_{ij}^{N}}L_{cls}\left(y_{ij}, \hat{y}_{ijk}^N\right)\\
    \mathcal{L}_{reg}({y_{ij}}) &=  
    \sum_{k=1}^{K_{ij}^{P}}L_{reg}\left(y_{ij}, \hat{y}_{ijk}^P\right)
    \label{eq:CE}
\end{align}
% \begin{align}
% % \mathcal{L} &= \sum_{i=1}^{I}\left[\mathcal{L}_{cls}({G_{i}}) + \mathcal{L}_{reg}({G_{i}})\right]\\
% \mathcal{L}_{cls}({G_{i})} &=
%      \sum_{j=1}^{J_{i}}L_{cls}\left(G_{i}, P_{ij}\right) + \sum_{k=1}^{K_{i}}L_{cls}\left(G_{i}, N_{ik}\right)
%     % \mathcal{L}_{reg}({G_{i}}) &=  
%     % \sum_{j=1}^{J_{i}}L_{reg}\left(G_{i}, P_{ij}\right)
%     \label{eq:CE}
% \end{align}
% \begin{align}
% % \mathcal{L} &= \sum_{i=1}^{I}\left[\mathcal{L}_{cls}({G_{i}}) + \mathcal{L}_{reg}({G_{i}})\right]\\
% % \mathcal{L}_{cls}({G_{i})} &=
% %      \sum_{j=1}^{J_{i}}L_{cls}\left(G_{i}, P_{ij}\right) + \sum_{k=1}^{K_{i}}L_{cls}\left(G_{i}, N_{ik}\right)\\
%     \mathcal{L}_{reg}({G_{i}}) &=  
%     \sum_{j=1}^{J_{i}}L_{reg}\left(G_{i}, P_{ij}\right)
%     \label{eq:CE}
% \end{align}
% $\hat{Y}_i = \{\hat{y}_{ik}\}_{k=1}^K$
where there are $K_{ij}^{P}$ positive segments $\hat{Y}_{ij}^{P}= \{\hat{y}_{ijk}^P\}_{k=1}^{K_{ij}^{P}}$
% $\hat{Y}_i = \{\hat{y}_{ik}\}_{k=1}^K$
, and $K_{ij}^{N}$ negative segments 
% $\hat{Y}_{ij}^{N}$
$\hat{Y}_{ij}^{N}= \{\hat{y}_{ijk}^{N}\}_{k=1}^{K_{ij}^{N}}$
for the \emph{j}-th ground-truth $y_{ij}$.
We use the focal loss \cite{Focalloss2018} for website classification, and the IoUT-based regression loss for segment localization.

{\bf Parameters update.}
For training, we adopt the standard supervised deep learning procedure.
The stochastic gradient descent (SGD) algorithm is used to update the model parameters  $\theta$ at every iteration $t$ as:
\begin{align}
    \theta(t+1)~=:\theta(t) - \alpha \nabla_{\theta} \mathcal{L}(\mathcal{D}_{train}^{batch};\theta)
\end{align}
where $\mathcal{L}$ denotes the training loss of the current mini-batch multi-traces with their ground-truths $\mathcal{D}_{train}^{batch}$.

% The above training process is no different between detecting the ground-truths and their clean ones.

\subsection{Model Testing}
In model test (attacking mode), 
% most of the proposal segments cannot be directly viewed as possible monitored traces.
among the $n$ candidate segments predicted at each cell segment, there is at most one correct monitored trace.
% but $K$ candidate segments are proposed.
Hence, a filtering process is needed.
% we should discard the impossible ones.
%
% {\bf Filtering. }
We take two steps to filter out low quality proposal segments. 
(1) We first filter out those segments with maximum website predicted probability lower than a predefined threshold value, i.e., low confident segments;
(2) For each remaining proposal segment, we then filter out those significantly overlapped and lower confidence segments. We use IoUT metric for overlapping.
% value is high than a predefined threshold value)
% whose top website predicted probability is lower than its.

% The above attack process is also no difference between the detection of ground-truths and their clean ones.
\subsection{Efficiency Optimization}
% The above architecture can give us an effective WFD model, but it could be further optimized.
For better efficiency, we further introduce three key designs:
compact data representation, lightweight feature extractor, and 2-staged training. 

\subsubsection{Compact data representation}
The raw traffic data of a multi-tab trace can be one or more of these types: time or directional representation of its {\bf \em cell} sequence, time or length representation of its {\bf \em burst} sequence.
For robustness against the network dynamics, we do not consider the time representation.
% For simplification, we only choose one sequence whose selection would be specified later.
\begin{myDef}
\label{cell}
{\bf Cell.} Unlike packets, a cell is the basic data unit of Tor, whose length is the same 512 bytes in the TLS records, denoted as ${c}$ = (${t}$, ${d}$) where ${t}$ is the cell's timestamp and ${d}$ is the cell's direction, $\pm1$.
${d}=+1$ means the cell is outgoing from the user to the website, and ${d}=-1$ means the cell is incoming from the website back to the user.
\end{myDef}
\begin{myDef}
\label{burst}
{\bf Burst.}  A burst is the maximum accumulation of a  consecutive cell sequence of  $(c_{1},c_{2}, \cdots)$ in the same direction, denoted as $b$ = (${t}$, ${l}$) where ${l} = \pm{|b|}$, $|.|$ is the length operation of the sequence, and $|b|$ is the number of cells in the burst.
\end{myDef}

% A compact data representation of the multi-tab trace is the first thing to consider.
Although the directional representation of cell sequence is robust and good to represent a single-tab trace, its sequence would be very long.
% as the input of \MT~model.
% to train  or test while representing a multi-tab trace.
In contrast, the lengthy representation of burst sequence is more compact, typically about one-tenth of its cell sequence, as verified by the statistics on the single-tab trace dataset DS-19 \cite{Gong2020}.
% Especially, using this compact data representation, we find that 1D ResNet series could easily achieve the start-of-art \ST~performance.
Compared with cell representation, burst reduces the training time, accelerating the WF attack process. % and makes itself more attractive.
% Hence, we use the lengthy representation of burst sequence to represent the multi-tab trace.

\subsubsection{Lightweight feature extractor}
Using burst sequences enables us to choose lightweight ResNets \cite{He2016},
with representative models 1D-ResNet18, 1D-ResNet34, and 1D-ResNet50.
In \ST, these models are similarly performing.
However, we find that 1D-ResNet18 performs best among these in \MT.

% Considering burst sequence unused for \WFA~before, 
% We choose the best one from the noticeable ResNet series \cite{He2016}, including 1D-ResNet18, 1D-ResNet34, and 1D-ResNet50.
% Although their performance are almost no difference for \ST, we find that the 1D-ResNet18 model performs better than the other candidates while using the pre-trained model as the frozen feature extractor for the \MT.
% Furthermore, its parameters are fewer and the inference speed is higher.

\subsubsection{2-staged training}
As discussed in Section \ref{sec:remarks}, the bootstrap time for \WFA~is crucial.
In particular, the model training time takes a big portion.
We propose a 2-stage training pipeline for acceleration.
In first stage, we ptr-train our feature extractor on a large \ST~dataset.
This is based on our observation that clean trace segments of multi-tab traces are similar to single-tab traces.
In second stage, we freeze the feature extractor and only train the remaining part (the scale encoder and the two-head predictor) of our model on the \MT~training set.
This can reduce the whole training time due to less update on the feature extractor.
%

% one, hence we can represent them in the same way.
% More specifically, we can represent the multi-tab trace with a pre-trained \ST~feature extractor.
% For the WFD model, if its feature extractor is pre-trained beforehand and then frozen, the whole training could be accelerated needless of the significant time-consuming parameter updating.
% With the saving time, the successive scale encoder can fully focus on the scale information, and the two-head predictor only pays attention to their predictions.
% Hence, we pre-train the feature extractor on a large-scale single-tab trace dataset and freeze it for later training of the WFD model.

\section{Experiments}
\subsection{Datasets and Protocols}
%
% We introduce two multi-tab datasets, one for normal training and test, 
% and the other for testing the generality and practicality.
% The GLUE dataset is used to test the performance of our WFD model against the latest defense.

We use two existing single-tab trace datasets.
% AWF \cite{Rimmer2018} for pre-training our WFD model, DS-19 \cite{Gong2020} for generating a multi-tab dataset.
%
(1) \textbf{AWF dataset} \cite{Rimmer2018}:
This dataset AWF$^\text{900}$ contains a total of 900 monitored target websites, each with 2,500 raw feature traces. 
All 900 websites are divided into a split of 576/144/180 for training ({AWF$^\text{tr}$}), validation ({AWF$^\text{va}$}) and test ({AWF$^\text{te}$}), respectively.
We use this dataset to pre-train the feature extractor of our WFD model.
(2) \textbf{DS-19 dataset} \cite{Gong2020}:
% This dataset is fresh.
{There are} 10,000 raw feature samples from 100 monitored websites.
Besides, another 10,000 raw feature unmonitored samples are collected. 
We divide DS-19 dataset into two disjoint parts: DS-19$_\text{ATTACKTRAIN}$ (11,000 samples including 55 samples per monitored website and 5500 non-monitored samples) and DS-19$_\text{EVALUATION}$ (9,000 samples including 45 samples per monitored website and 4500 non-monitored samples). 
We use DS-19$_\text{ATTACKTRAIN}$ to generate the multi-tab training dataset for \MT; DS-19$_\text{EVALUATION}$ to synthesize the multi-tab test dataset.

We create two new synthetic \MT~datasets as there are no options in the literature, followed by a GLUE dataset.
(1) {\bf DS-19$_{mt}$: }
In this \MT~dataset, each simulated multi-tab trace is specified as a $\ell$-tab trace consisting of $\ell$ single-tab traces with overlap between neighbours in align with realistic multi-tab visiting patterns.
There are three overlapping positions: tail (clean front segment), front (clean tail segment), both ends (clean middle segment).
For each case, we set six overlapping percentages from 10\% to 60\%  with a step of 10\%.
As a result, there are a total of 18 different overlapping types.
With the single-tab traces from DS-19$_\text{ATTACKTRAIN}$, we generate  $\lfloor9900\times 18 /\ell\rfloor$ $\ell$-tab traces for each $\ell$ $\in$ $\left[2, 5, 8, 11, 14, 17\right]$ and the overlapping types are randomly selected.
With the base rate $r=10$, there are $900\times 18$ single monitored traces and $9000\times 18$ single unmonitored traces for every $\ell$ case.
% no matter for which $\ell$-tab traces.
All these generated traces with different $\ell$ values
are used as \MT~training data.
% After generation, all these $\ell$-tab traces are used as a whole for the model training.
We generate the \MT~test data in the same way using 
DS-19$_\text{EVALUATION}$.
% The generation of test data is the same as the training data, except the single-tab traces are from the DS-19$_\text{EVALUATION}$.
% This multi-tab dataset DS-19$_{mt}$ is mainly set to compare the attack performance of our WFD model and the start-of-the-art CDSB series.
%

(2) {\bf DS-19$_{mt}^{hard}$: }
This is a harder \MT~dataset.
For the test data, we generate $\lfloor900 \times (r + 1) \times 18 /\ell\rfloor$ $\ell$-$r$-traces for every combination of $\ell$ $\in$ $\left[5, 7, 9, 11, 13, 15, 17\right]$ and $r$ $\in$ $\left[1,3,5,7,9,11\right]$.
We use the same overlapping setting.
%
% About the overlapping, we follow the above setting.
We create two training situations: %for the training, 
(i) normal training data size as in typical benchmarks, %and test distribution,
(ii) small training data size.
%
% one for normal training whose generation is similar as the test data except the source of single-tab traces, 
% (2) 
% another for the light training.
% For the former, this multi-tab dataset DS-19$_{mt}^{hard}$ is mainly set to study the generality and practicality of our WFD model.
%
For the latter, we set $\ell =3$ and the base rate $r=1$,
giving only 10800 $\ell$=3-$r$=1-traces for training.

% For the light training data, we just set 
% the $\ell$ as 3, and the base rate $r$ as 1.
% Hence, there are 10800 $\ell$=3-$r$=1-traces, only a very little fraction 
% % (smaller than 1/42) 
% of the normal training data. 
% With the normal/light training data, this multi-tab dataset DS-19$_{mt}^{hard}$ is mainly set to study the generality and practicality of our WFD model.

(3) {\bf DS-19$_{GLUE}$: }
This is a GLUE \MT~dataset~\cite{Gong2020}. Each GLUE multi-tab trace is a $\ell$-tab traces with $\ell$ single-tab traces glued by the GLUE defense.
The same setup of $\ell$ as DS-19$_{mt}$ is used, 
% The $\ell$-tab trace setting of DS-19$_{GLUE}$ is the same as the above DS-19$_{mt}$, 
except no single-tab traces.
% with 16 $\ell$ numbers from 2 to 17.
For the FRONT setting, we follow Zero-delay's setting (N$_{s}$ = N$_{c}$ = 1100, W$_{min}$ = 10 s, W$_{max}$ = 15 s) \cite{Gong2020}.
This dataset is designed for testing our WFD model against the latest defense GLUE \cite{Gong2020}.

In the following experiments, we consider
the more realistic \emph{open-world} scenario.

% evaluate our WFD model on three standard WF attack datasets in \emph{open-world} scenario.

\renewcommand{\arraystretch}{1.5} %控制行高
\begin{table} [th] %[!tp]
  \centering
  \setlength{\tabcolsep}{4pt}
  \caption{Overall results of our WFD model and the latest CDSB series (\emph{k}NN, CUMUL, \emph{k}FP and DF)  on DS-19$_{mt}$.
  Metrics: precision (\%), recall (\%), mAP (\%), mAP$_{.50}$ (\%), and mAP$_{.75}$ (\%). 
  }
  \label{tab:overlap_all}
  {
    \scalebox{0.9}{
    \begin{tabular}{c|cccccccc}
    \toprule
    Method & Precision & Recall & mAP & mAP$_{.50}$ & mAP$_{.75}$ \cr
    \midrule
    CDSB+\emph{k}NN   & 13.97 & 6.71  & 4.88 &5.90 &4.81\cr
    CDSB+CUMUL        & 1.28  & 7.25  & 1.70 &2.12 &1.68\cr
    CDSB+\emph{k}-FP  & 11.13 & 7.44  & 5.78 &7.52 &5.64\cr
    CDSB+DF            & 2.73 & 10.82& 7.83 &11.26&7.42\cr
    \hline
    {\bf WFD (Ours)} & {\bf 84.09} & {\bf 62.64} & {\bf 70.49} &{\bf 83.38}&{\bf 76.77}\cr
    % {\bf WFD-clean (Ours)} & {\bf 81.44} & {\bf 74.50} & {\bf 73.42} &{\bf 90.31}&{\bf 83.67}\cr
    \bottomrule
    \end{tabular}
    }}
\end{table}
\begin{table} [th] %[!tp]
  \centering
  \setlength{\tabcolsep}{10pt}
  \caption{Evaluating the training time and attack speed
%   , evaluation results of our WFD model and the latest CDSB series (\emph{k}NN, CUMUL, \emph{k}FP, and DF)  
  on DS-19$_{mt}$.
  Metrics: train time in minutes, attack speed in MBps.
  }
  \label{tab:overlap_speed}
  {
    \scalebox{0.9}{
    \begin{tabular}{c|cc}
    \toprule
    Method & train-time & test-speed  \cr
    \midrule
    % CDSB+\emph{k}NN & 133 & 12:57:41  \cr
    % CDSB+CUMUL & 167 & 1:43:05  \cr
    % CDSB+\emph{k}-FP  & 90 & 10:02:37\cr
    % CDSB+DF  & 152 & 10:19:33\cr
    % \hline
    % Our (WF detection)  & 54 & 33:32 \cr
        CDSB+\emph{k}NN & 133 & 52.10  \cr
    CDSB+CUMUL & 167 & 393.06  \cr
    CDSB+\emph{k}-FP  & 90 & 67.24\cr
    CDSB+DF  & 152 & 65.40\cr
    \hline
    {\bf WFD (Ours)}  & {\bf 54} & {\bf 1208.30} \cr
    \bottomrule
    \end{tabular}
    }}
\end{table}

\subsection{Implementation Details}
\keypoint{Benchmarks. } We compare the state-of-art \MT~model \cite{Gong2020} originally proposed for tackling the GLUE defense. In particular, Gong et al. \cite{Gong2020} 
took a two-stage strategy of split decision and split finding in
the Coarse-Decided Score-Based (CDSB) framework. After single-tab traces are obtained via splitting, \emph{k}NN, CUMUL, \emph{k}FP and DF are then used to classify the single-tab traces.  

\keypoint{Raw input. } For the training and test of CDSB series, the original cell sequence (both directional and time information) is used to represent trace samples.
For our WFD model, the more compact burst sequence (lengthy information) is used.
\begin{figure}%[h!]
    \centering
    \includegraphics[scale=1.11]{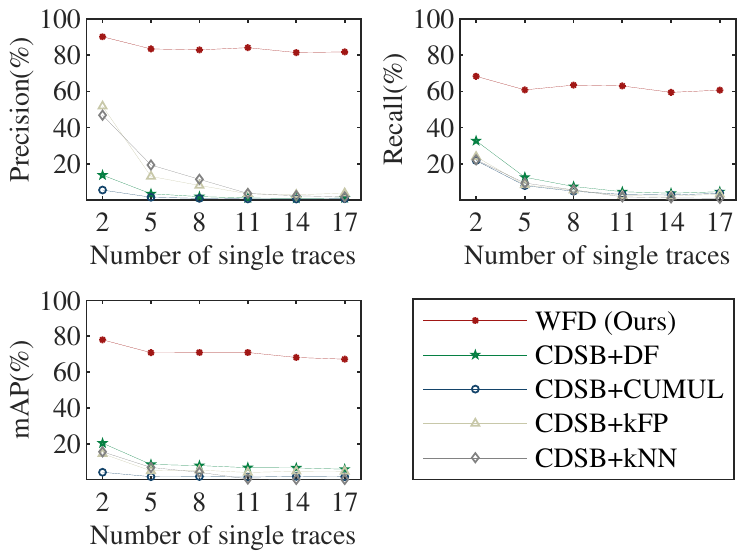}
    \caption{
    Evaluating the effect of single-tab trace number ($\ell$) in each multi-tab trace 
    % with our WFD on
    % $\ell$-tab traces evaluation results of our WFD model and the latest CDSB series (\emph{k}NN, CUMUL, \emph{k}FP and DF)  
    on DS-19$_{mt}$.
  Metrics: precision (\%), recall (\%), and mAP (\%).
}
\label{fig:overlap_l}
\end{figure}

\keypoint{WFD model.}
We choose 1D ResNet18 as feature extractor.
For the scale encoder, we set 4 dilated blocks with the dilated rates of [2,4,6,8].
% For the two-head predictor, the number layers of regression head is x, and the classification head's is x.

\keypoint{WFD Training.}
For training WFD, we pre-train the feature extractor on the AWF dataset in the first stage, freeze the pre-trained feature extractor when training the remaining parts in the second stage.
We adopt the SGD optimizer with a learning rate of 0.12 and
a batch-size of 48.
We train a total of 6000 iterations.
We use an NVIDIA 3090 GPU for all the experiments.

\subsection{Evaluation on Realistic \MT}
\keypoint{Setting. }
In this experiment, we use the multi-tab dataset DS-19$_{mt}$ to train and test our WFD model and the CDSB series.
For the training of CDSB, we divide the single-tab trace dataset DS-19$_{TRAIN}$ into two parts: 2000 samples for SPLITTRAIN and another 9000 samples for ATTACKTRAIN.
This way, we use the same original training data for 
CDSB and our WFD for fair comparison.
% we could keep the training data setting of CDSB with ours, whose training data is also generated from  DS-19$_{TRAIN}$. 
We use the officially released code of CDSB in our experiments.
We use both the accuracy metrics (precision, recall, and mAP) and efficiency metrics (training time in minutes, attack speed in MBps).

\begin{figure}%[h!]
    \centering
    \includegraphics[scale=1.1]{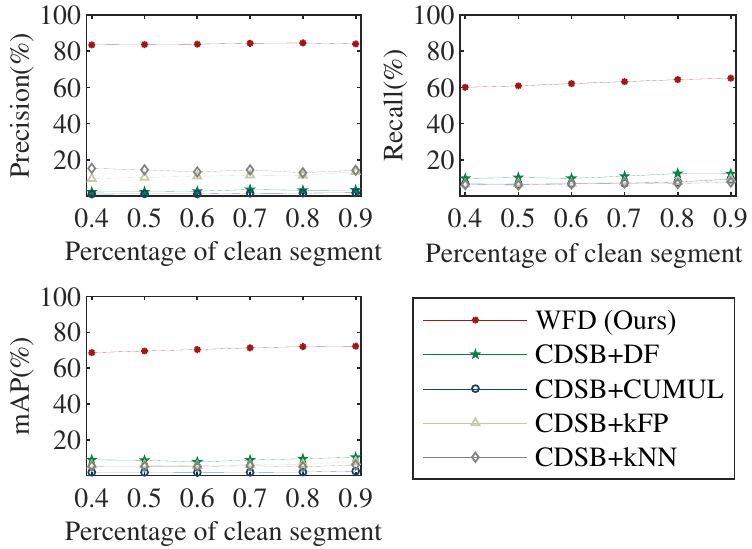}
    \caption{
    Evaluating the effect of the percentage of clean trace segments
    % For different percentages of the clean monitored segment in its full trace, evaluation results of our WFD model and the latest CDSB series (\emph{k}NN, CUMUL, \emph{k}FP, and DF)  
    on DS-19$_{mt}$.
  Metrics: precision (\%), recall (\%), and mAP (\%).
}
\label{fig:overlap_p}
\end{figure}

\keypoint{Results. }
The comparative results are reported in Tables \ref{tab:overlap_all}, \ref{tab:overlap_speed}, and Figures \ref{fig:overlap_l},  \ref{fig:overlap_pos}, \ref{fig:overlap_range}, \ref{fig:overlap_p}.
We have the following observations and discussions.
\begin{enumerate}
\item 
In all accuracy metrics (mAP, precision, and recall), our WFD model is consistently superior over the state-of-the-art \MT~alternatives, the CDSB series (kNN, CUMUL, kFP, and DF). 
Note that, the most significant margin is achieved in the most difficult $\ell=17$ case, with a minimum edge of 61.49\% of mAP, 78.23\% of precision, 56.23\% of recall, over the start-of-the-art alternative method.
This implies the robust stability of our WFD across varying length multi-tab traces. 
\begin{figure}%[h!]
    \centering
    \includegraphics[scale=1.1]{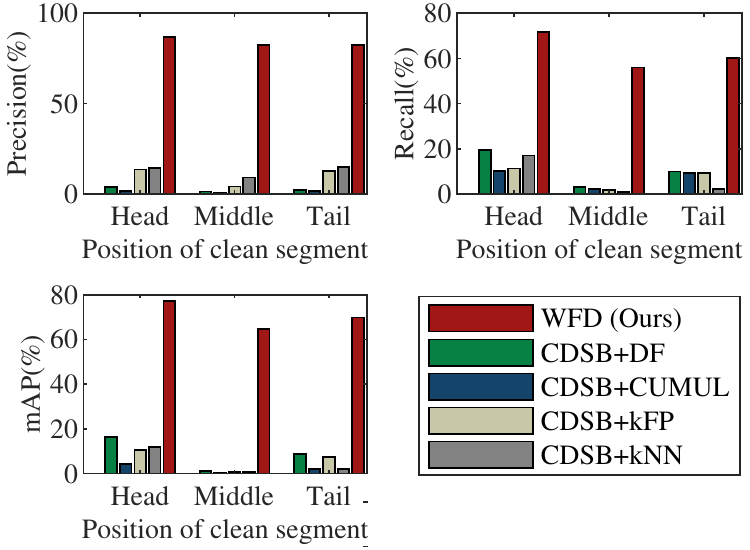}
    \caption{
    Evaluating different positions (head, tail, middle) of clean trace segments
    % in its full trace, evaluation results of our WFD model and the latest CDSB series (\emph{k}NN, CUMUL, \emph{k}FP, and DF)  
    on DS-19$_{mt}$.
  Metrics: precision (\%), recall (\%), and mAP (\%).
}
\label{fig:overlap_pos}
\end{figure}
\item 
In AP metrics, AP$_{50}$ results are always better than AP$_{75}$.
This is as expected, as higher IoUT means a more strict metric,
and it is often harder to find a more accurate location.
% For AP metrics, AP$_{50}$ results are always the best, followed by AP$_{75}$ results, and the last is mAP results.
% This means that it is hard to optimize location and classification simultaneously.
% When we pursue the advance of WFD model on the classification metrics, its location performance is hardly advanced at the same time.
\item
As shown in Figure \ref{fig:overlap_pos}, the position of clean segment (front, middle or tail) is important for detection.
While the clean segment is in the front, the detection performance is best.
These results are consistent with the fact that the front of a trace is most informative, which motivates the FRONT defense \cite{Gong2020}.
Additionally, the detection performance would be better if the clean segment is located in the tail as compared to in the middle part.
\begin{figure}%[h!]
    \centering
    \includegraphics[scale=1.09]{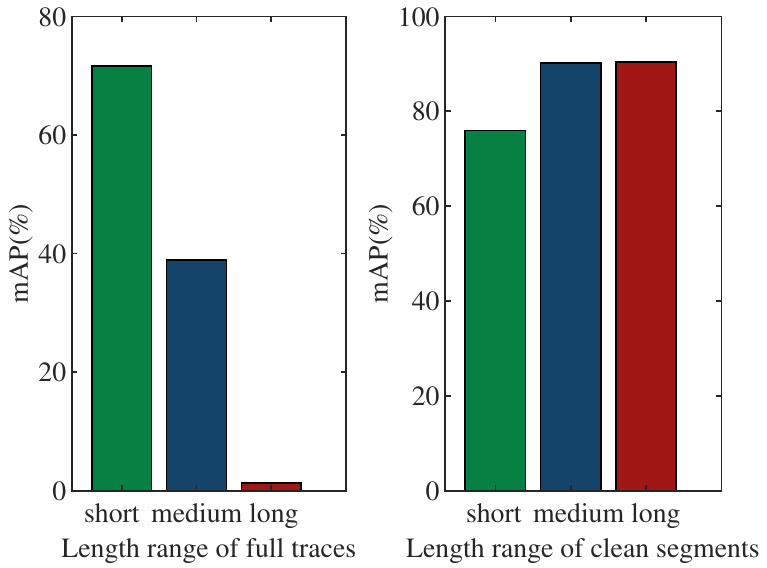}
    \caption{
    % In the multi-tab traces, for the lengths of full ground-truth traces and their clean segments, evaluation results of our WFD model. %Regarding full traces, 
    Evaluating the effect of trace length and clean segment length 
    with our WFD model on DS-19$_{mt}$ dataset. 
    The range of trace lengths: [0-1000] for short ones, 
    [1000-4000] for medium ones, and [4000-9000] for long ones. 
    The ranges of clean segment lengths: [0-500] for short ones, [500-1000] for medium ones, and [1000-4000] for long ones.
    Metrics: mAP (\%).
}
\label{fig:overlap_range}
\end{figure}
\item
As shown in Figure \ref{fig:overlap_range}, the length of a clean trace is vital for detection, but not the entire length. % of its ground-truth.
The longer the clean segment is, the easier the trace would be detected.
Detecting the test traces with short clean segments is harder than 
those with long clean segments, with mAP increasing from 75.94\% to 90.48\%.
% Compared to the test traces with short clean segments, the detection of those whose clean segments are long is more accurate, whose mAP value increases from 75.94\% to 90.48\%.
However, an opposite situation happens 
on the relation between the detection result and the trace's entire length.
The longer the entire trace is, it would be more challenging to be detected.
This is because a long trace is more easily overlapped with others which
challenges the detection model significantly.
% a main contributor of its full length is the overlapping parts which cannot help the detection.
These results suggest that the detection performance of entire traces depends on the length of clean segments.

% the data pattern concealed in a shorter clean trace segment is harder to extract compared to the longer one.

    \item 
    Except the absolute length of clean segments,
    their proportion is another vital factor for detection.
% Besides the impact of the whole length of  ground-truth's clean segment, its percentage is also vital  to detection.
As shown in Figure \ref{fig:overlap_p}, when the percentage of clean segment 
increases, the detection becomes easier, with mAP increased from 68.67\% to 72.28\%.
% in the ground-truth becomes more, the detection of its ground-truth becomes easier, from 68.67\% to 72.28\% on mAP. 
    \item
For the training time, our WFD model consumes the fewest time (54 minutes), 
almost double the training speed of the previous fastest model CDSB+\emph{k}-FP.
% around half of the CDSB+\emph{k}-FP's, the previous fastest training model.
Further, our attack speed is also the best,
about 20 times faster than the previous most accurate model CDSB+DF.

%
% Except for the acceleration of training, our attack speed is also the highest, about 20 times faster than the previous most accurate model CDSB+DF.
On an ordinary NVIDIA 3090 GPU, WFD achieves a \MT~speed of 1208.30 MBps.
In case of minimum 1 second webpage loading time and average 2.27 MB (average 4441 cells per single-tab trace in DS-19) traffic flow per webpage, our WFD can simultaneously attack at least 532 (e.g., 1208.3/ 2.27 = 532) users at real time.
For the mean 16 to 19 seconds to load web pages over Tor browser \cite{WangBlast2020}, we can attack simultaneously 8516 to 10113 users at real time.
This implies that our WFD has the potential for large-scale real-time WF attack.

\end{enumerate}

\begin{table} [th] %[!tp]
  \centering
  \setlength{\tabcolsep}{10pt}
  \caption{
  Evaluating the training time and accuracy of our WFD model 
  with different training data sizes
%   trained only with $\ell$=3-$r$=1-traces (Small trained) 
  on DS-19$_{mt}^{hard}$.
%   With respect to training with different $\ell$-$r$-traces (Normal trained), the training time and evaluation results of our WFD model trained only with $\ell$=3-$r$=1-traces (Small trained) on DS-19$_{mt}^{hard}$.
  Metrics: train time in minutes, attack speed in MBps.
%   Metrics: minutes and mAP(\%).
  }
  \label{tab:hard_tm}
  {
    \scalebox{0.9}{
    \begin{tabular}{c|cc}
    \toprule
    Training data size &Train-time & mAP \cr
    \midrule
    Normal  & 105 & {\bf 69.33} \cr
    \hline
    Small  & {\bf 64} & 60.55 \cr
    \bottomrule
    \end{tabular}
    }}
\end{table}

\subsection{Evaluation on Generality and Practicality}
\keypoint{Setting. }
This experiment is conducted on DS-19$_{mt}^{hard}$.
We consider two training cases:
normal and small training data sizes.
% our WFD model with two types: one is the normal trained model, another is the light model that is trained only with a small number of $\ell$=3-$r$=1-traces.
We test the two trained models on a large number of $\ell$-$r$-traces generated with different single-tab traces numbers $l$ and different base rates $r$.
We use mAP as accuracy metrics.
For efficiency metrics, we use minutes for training time and MBps for attack speed.
\begin{figure}%[h!]
    \centering
    \includegraphics[scale=1.1]{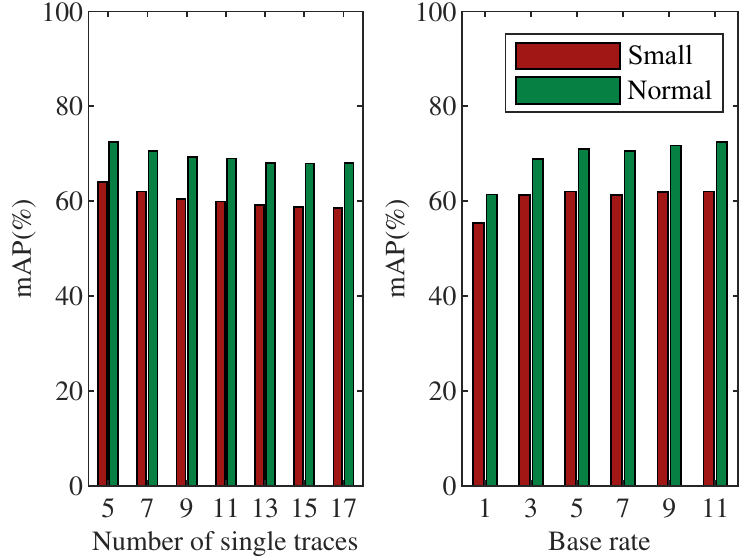}
    \caption{
    Evaluating the effect of training data size with  our WFD model
    % With respect to training with different $\ell$-$r$-traces (Normal trained), evaluation results of our WFD model trained only with $\ell$=3-$r$=1-traces (Small trained) 
    on DS-19$_{mt}^{hard}$.
    Metrics: mAP (\%).
}
\label{fig:hard_traces}
\end{figure}

\keypoint{Results. }
The comparative results are reported in Table \ref{tab:hard_tm} and Figure \ref{fig:hard_traces}.
We have the following observations.
\begin{enumerate}
\item
Given the normal training data size and 
testing with a different base rate ($r$ $\in$ $\left[1,3,5,7,9,11\right]$)
% WFD model, 
WFD still achieves a similar mAP
as compared to in the same base rate $r$=10,
69.33\% vs. 70.49\%.
%
% performance (69.33\% of mAP) testing with a changing base rate ($r$ $\in$ $\left[1,3,5,7,9,11\right]$) is almost not different from the performance (70.49\% of mAP) training and testing with a fixed base rate $r$=10.
This implies our WFD model is robust to the base rate change.

\item 
Testing with different $\ell$-tab traces,
% with a different number $\ell$ of single-tab traces, 
our WFD trained with only $\ell$=3 single-tab traces still performs well 
as in case of normal training data size, with mAP varying from 60.55\% to 69.33\%.
This suggests good generality of our WFD across varying numbers $\ell$ of single-tab traces in a single multi-tab trace.
Therefore, there is no tedious need for preparing a specific training data for every $\ell$. % which is  if there is an emergency.
% Another interesting thing is that 
On the contrary,
the competitors all degrades clearly with the increase of single-tab traces.
% , no matter for the Small trained model or normal trained one.
\item 
% Compared with the normal trained model, the Small trained WFD model 
Encouragingly, even with small training data, WFD 
can still perform well, giving at least 55.46\% mAP,
vs 62.02\% (the least mAP for normal training data size).
This suggests our WFD model has a good  generality and practicality.
This ability is important for realistic \MT~as the base rate and the number of single-tab traces would frequently change.
Interestingly, the performance is even better when the base rate $r$ increases.
This result is favorable for attacking the cases with significant base rates, often occurring in realistic environments.
    
    \item
For the bootstrap time, given small training data, the training only needs 64 minutes, vs 105 minutes in the normal counterpart.
% Furthermore, there is still the saved time of data preparation for the collection of realistic multi-tab traces.
This suggests that with WFD, we do not need to take much time on training data collection.
% This suggests that our WFD model can bootstrap quickly, needless of a long preparation.
\end{enumerate}

\subsection{Evaluation on GLUE Defense}
\label{subsec:glue_eva}
\keypoint{Setting. } 
In this experiment, 
we use the DS-19$_{GLUE}$ dataset for our WFD model and the CDSB series.
% the above dataset DS-19$_{mt}$ except its overlapping.
It is noteworthy that the CDSB needs two trained WF models: 
(1) a ``noisy model'' 
% for FRONT and another "clean model".
trained on traces with FRONT noise to classify the first single-tab trace defended with FRONT; 
(2) a ``clean model'' trained without FRONT noise to classify the other single-tab traces.
In contrast, we train the WFD model in a single manner by viewing the traces of each monitored website with FRONT noise as the original website's.
% Our other training setting is kept unchanged.
We use precision, recall, and mAP as the accuracy metrics. 
\renewcommand{\arraystretch}{1.5} %控制行高
\begin{table} [th] %[!tp]
  \centering
  \setlength{\tabcolsep}{15pt}
  \caption{Overall results of our WFD model and the latest CDSB series (\emph{k}NN, CUMUL, \emph{k}FP and DF) on DS-19$_{GLUE}$.
  Metrics: precision (\%), recall (\%) and mAP (\%).
  }
  \setlength{\tabcolsep}{12pt}
  \label{tab:glue_t}
  {
    \scalebox{0.9}{
    \begin{tabular}{c|cccc}
    \toprule
    Method & Precision & Recall & mAP
    \cr
    CDSB+\emph{k}NN &3.89&3.93&2.76\cr
    CDSB+CUMUL & 1.25&6.95&1.91\cr
    CDSB+\emph{k}-FP  &3.76&4.67&4.25\cr
    CDSB+DF  &2.89&13.79&15.81 \cr
    \hline
    {\bf WFD (Ours)}  & {\bf 88.32}&{\bf 82.98}&{\bf 80.56} \cr
    \bottomrule
    \end{tabular}
    }}

\end{table}
\begin{figure}%[h!]
    \centering
    \includegraphics[scale=1.1]{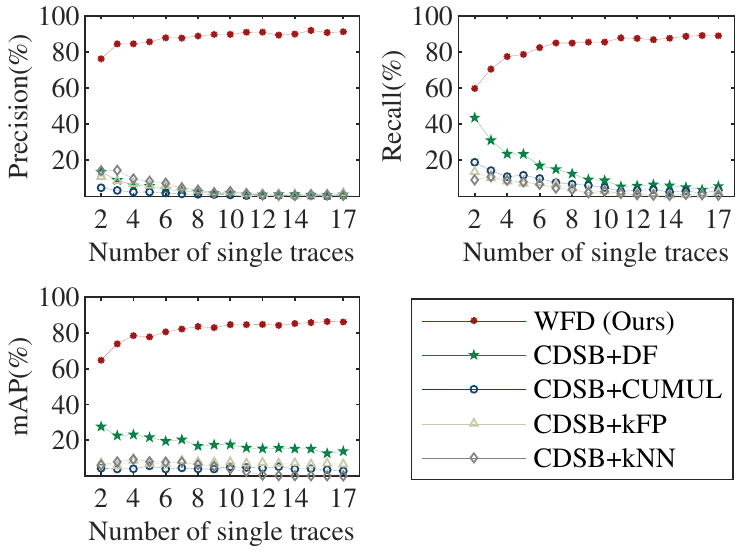}
    \caption{
    Evaluating the effect of single-tab trace number ($\ell$) in a single multi-tab trace
    % $\ell$-tab traces evaluation results of our WFD model and the latest CDSB series (\emph{k}NN, CUMUL, \emph{k}FP and DF)  
    on DS-19$_{GLUE}$.
  Metrics: precision (\%), recall (\%), and mAP (\%).
}
\label{fig:glue_4}
\end{figure}

\keypoint{Results. }
The comparative results are shown
in Table \ref{tab:glue_t}
and Figure \ref{fig:glue_4}.
We make the following observations:
\begin{enumerate}
    \item Although the single-tab traces are defended with the GLUE defense,
our WFD model surpasses CDSB series (kNN, CUMUL, kFP, and DF) consistently, with a minimum margin of 84.5\% in precision, 69.2\% in recall, and 64.75\% in mAP.
\item 
In all the metrics, 
all \MT~models yield lower accuracy on multi-tab traces with overlapping
than on the GLUE-defended ones, consistently.
% are always inferior to on the GLUE-defended single-tab ones.
% for all multi-tab WF attack models, their performance on the multi-tab traces with overlapping are always inferior to on the GLUE-defended single-tab ones.
This suggests that realistic \MT~is more challenging than the GLUE-defended traces.
    \item 
With the increase of single-tab traces in the glued trace, the performance of CDSB series decreases.
This is because split decision and split finding by CDSB series depend heavily on the number $\ell$ of single-tab traces.
% However, our WFD model is immune to the number change of single traces.

    \item
For our WFD model, the performance is even better when the number $\ell$ of single-tab traces is increased.
This is because the percentage of FRONT-defended single-tab trace becomes small given more single-tab traces.
This shows that the FRONT defense is still powerful for our WFD model,
although strong results have been obtained 
% although we achieve quite well success 
on these glued ones.
\end{enumerate}

\section{Conclusion}
In this work, we investigate for the first time the realistic and more challenging \MT~problem. This eliminates the unrealistic assumptions of sequential single website visiting and manual trimming as made in conventional \ST~works.
Going beyond classifying manually trimmed single traces, 
we take a novel {\em detection} perspective. 
We further propose the first end-to-end {\bf \em Website Fingerprint Detection} (WFD) model, particularly designed for solving the \MT~problem.
This is inspired by the success of object detection in {\color{black}computer vision}.
Our model is different drastically from
existing alternative methods that consider trace localization
and classification independently, without the ability 
to optimize their compatibility.
We also introduce several designs to improve the model efficiency.
Extensive experiments have conducted to validate the significant superiority of our WFD over the start-of-the-art models in the \emph{open-world} setting with different overlapping cases,
% positions, different overlapping percentages, 
different numbers of single-tab traces, 
different base rates, with and without defense, with normal and small training data.

\bibliographystyle{IEEEtran}

\section{Appendix: Ablation Study}
\subsection{Data Representation}
We study the impact of original feature representation (the lengthy representation of burst sequence and the directional representation of cell sequence) on the performance of the \WFA~model.
\label{subsec:closed_world_exp}

\keypoint{Setting. } 
To simplify the experiment setting, we choose the single-tab trace attack in the \emph{open-world} scenario and use the famous DF \cite{Sirinam2018} as the WF attack model.
On the single-tab trace dataset DS-19 \cite{Gong2020}, we trained DF with the above two types of feature representation.
We use the officially released code of DF \cite{Sirinam2018} in our experiments and follow their default settings.
We use precision and recall as the performance metrics. 
\begin{table} [th] %[!tp]
  \centering
  \setlength{\tabcolsep}{10pt}
  \caption{
  With the lengthy representation of burst sequence and directional representation of cell sequence,
  evaluation Results of DF on DS-19.
  Metrics: precision (\%) and recall (\%).}
  \label{tab:abl_dr_pr}
  {
    \scalebox{0.9}{
    \begin{tabular}{c|cc}
    \toprule
    Type  & Precision & Recall \cr
    \midrule
    % Cell  & 0.9981515711645101 & 0.9574468085106383 \cr
    Cell Sequence  & {\bf 99.8} & {\bf 95.7}\cr
    \hline
    Burst Sequence  & 98.2 & 91.7\cr
    \bottomrule
    \end{tabular}
    }}
\end{table}
\begin{table} [th] %[!tp]
  \centering
  \setlength{\tabcolsep}{10pt}
  \caption{
  The average length of burst sequence and cell sequence of samples in DS-19, and the training time and test time of DF with each type of sequences padding to its max length with 0.
%   (4000 for burst sequence, 10000 for cell sequence). 
  Metrics: second.
  }
  \label{tab:abl_dr_time}
  {
    \scalebox{0.9}{
    \begin{tabular}{c|cc}
    \toprule
    Type & Cell Sequence & Burst Sequence \cr
    \midrule
    Average Length  & 5292 & {\bf 471}\cr
    Maximum Length & 10000 & {\bf 4000}\cr
    \hline
    Training time  & 161 & {\bf 70}\cr
    Test time &  55 & {\bf 14}\cr
    \bottomrule
    \end{tabular}
    }}
\end{table}

\keypoint{Results. }
The results of different methods are compared
in Table \ref{tab:abl_dr_pr}, and \ref{tab:abl_dr_time}.
We have the following observations:
\begin{enumerate}
    \item
    The precision result on burst sequences is almost the same as the results on cells.
    For recall, the evaluation result can also be acceptable, even with a small gap of 4\%. 
    There are two possible reasons:
    firstly, the data pattern in the burst sequence is not informative as the cell sequence due to its compression; 
    secondly, the DF model is designed for the cell sequence, and some specified models should be designed for the burst sequence.
    \item
    The average length of burst sequence is about one-tenth of the average length of cell sequence.
    Even for the maximum length, the burst sequence's is only about two-fifths of the cell sequence's.
    Undoubtedly, the burst sequence is a compact original feature sequence.
    \item
    The training time with burst sequence is almost two-fifths of the one  with cell sequence, consistent with the lengthy ratio of  burst sequence to  cell sequence.
    The test/attack time with burst sequence is almost one-fifth of the one with cell sequence, which is less than the ratio of training time (two-fifth) but more than the ratio of average length (about one-ten).
    The biggest advantage of burst representation is its high efficiency of both train and test, a bit lower but quite well classification performance with quite potential for improvement.
    Hence, we try to choose a suitable model for the burst sequence.
\end{enumerate}
\subsection{Lightweight Pre-trained Feature Extractor}
Since DF is specially designed for the cell sequence but not the burst sequence, we study the feature extraction ability of the 1D ResNet series (18, 34, 50) due to their wide applications.
We select the most lightweight 1D ResNet18 whose detection performance is even the best.
Using the pre-training and parameters-freezing technology, we save the bootstrap time cost to the maximum limit with the precondition of effectiveness.

\keypoint{Setting. } 
We pre-train these models on the dataset AWF$_{tr}$, and choose the best model performing on AWF$_{va}$, and finally report their 100-way 20-shot performance on AWF$_{te}$. 
The 100-way 20-shot performance means that, with 100 new websites of AWF$_{te}$, we freeze the feature extractor and only fine-tune its classifier with only 20 samples per website, and test the fine-tuned model in the closed-world scenario in which another 15 samples per website are randomly chosen.
Except for the input of a 512-length burst sequence, we follow the same training and test setting of Chen \cite{CHEN2021cn}, with the logistic regression classifier.
About the baseline TLFA \cite{CHEN2021cn}, we choose the 5000-length cell sequence as input, DF \cite{Sirinam2018} as the feature extractor, and the logistic regression as the classifier. 
We use accuracy as the performance metric.
For the WF detection evaluation of different feature extractors on DS-19$_{mt}$, we keep its setting the same as section \uppercase\expandafter{\romannumeral7}, but only training with 2000 iterations.

\begin{table} [th] %[!tp]
  \centering
  \setlength{\tabcolsep}{7pt}
  \caption{
  With 20 training samples  per new website, evaluation results of TLFA (DF, cell sequence), 1D-ResNet series (burst sequence) on AWF.
  Metrics: accuracy (\%).
  }
  \label{tab:abl_fe_pre_test_acc}
  {
    \scalebox{0.9}{
    \begin{tabular}{c|cccc}
    \toprule
    Method & TLFA &ResNet18 &  ResNet34 & ResNet50\cr
    \midrule
    Accuracy & {\bf 98.3} & 98.1 & 98.3 & 98.0\cr
    \bottomrule
    \end{tabular}
    }}
\end{table}
\begin{table} [th] %[!tp]
  \centering
  \setlength{\tabcolsep}{8pt}
  \caption{
  The training time per epoch
  TLFA (DF, cell sequence), 1D-ResNet series (burst sequence) on AWF
  Metrics: second.
  }
  \label{tab:abl_fe_pre_train_time}
  {
    \scalebox{0.9}{
    \begin{tabular}{c|cccc}
    \toprule
    Method & TLFA &ResNet18 &  ResNet34 & ResNet50\cr
    \midrule
    Training time&  211.35& {\bf 93.31} & 177.13 & 462.05\cr
    \bottomrule
    \end{tabular}
    }}
\end{table}
\begin{table} [th] %[!tp]
  \centering
  \setlength{\tabcolsep}{10pt}
  \caption{Evaluation results of WFD model with different parameter-frozen feature extractors: 1D-ResNet 18, 34, and 50 on DS-19$_{mt}$.
  Metrics: mAP (\%).
  }
  \label{tab:abl_fe_pre_test_det}
  {
    \scalebox{0.9}{
    \begin{tabular}{c|ccc}
    \toprule
    Method & ResNet 18 &  ResNet 34 & ResNet 50\cr
    \midrule
    % mAP  & 59.009 & 57.349 & 51.228\cr
        mAP  & {\bf 59.01} & 57.35 & 51.23\cr
    \bottomrule
    \end{tabular}
    }}
\end{table}
\begin{table} [th] %[!tp]
  \centering
  \setlength{\tabcolsep}{10pt}
  \caption{Compared with training from scratch, the evaluation results of the WFD model with parameters-frozen pre-trained feature extractors: 1D ResNet18 on DS-19$_{mt}$.
  Metrics: mAP(\%).
  }
  \label{tab:abl_fe_pre_test_scratch}
  {
    \scalebox{0.9}{
    \begin{tabular}{c|cc}
    \toprule
    Type &  From-scratch & Pre-trained\cr
    % mAP  & 20.523 & 59.009\cr
        mAP  & 20.52 & {\bf 59.01}\cr
        % \midrule
        % training time & xx & xx\cr
    \bottomrule
    \end{tabular}
    }}
\end{table}
\keypoint{Results. }
The comparative results are reported in Table \ref{tab:abl_fe_pre_test_acc},
\ref{tab:abl_fe_pre_train_time},
\ref{tab:abl_fe_pre_test_det}, and \ref{tab:abl_fe_pre_test_scratch}.
We have the following observations and discussions.
\begin{enumerate}
\item
For the few-shot \WFA~of these three pre-trained 1D ResNet models, their 100-way 20-shot accuracy is almost no different from each other, about 98\%, which is almost the same as TLFA(DF, cell).
From the view of the single-tab WF attack's effectiveness, 1D ResNet series with the lengthy representation of burst sequence as input are comparable to TLFA(DF, cell) with the directional representation of cell sequence.
\item
From the view of pre-training time, the 1D ResNet18's is obviously less than the three others, about the half of ResNet34's and TLFA(DF,cell)'s,  and one-fifth of ResNet50's.
This suggests the pre-train time superiority of 1D ResNet18.
\item
The parameters amount (1,575,296) of 1D ResNet18 is almost not different from the DF's (1,438,784) in order of magnitude.
Moreover, the burst sequence length is only about one-tenth of its corresponding cell sequence length.
Hence, the combination of 1D ResNet18 and burst sequence is a best choice not only for \ST~but also for \MT.
\item
To save the training time of our WFD model, we freeze the parameters of these three pre-trained 1D ResNet models and use them as the feature extractor.
With fewer parameters, the 1D ResNet18 performs better than its two competitors.
Compared with the 1D ResNet18, the frozen parameters of 1D ResNet34 and 1D ResNet50 are more, the data patterns extracted in it are more complex, it would be more difficult for the scale encoder and predictor to train with them.
This results support us to choice the parameter-frozen pre-trained 1D ResNet18 as the feature extractor of our WFD model. 
\item
Compared to training from scratch, with the same training interations, the WFD model with parameter-frozen pre-trained 1D ResNet18 as feature extractor performs obviously better, with a significant  margin of 38.49 mAP.
Maybe more training time is needed for the WFD model training from scratch to achieve the same performance as with the pre-trained feature extractor.
This verifies the necessity and time-saving of pre-training.
\end{enumerate}

\subsection{Single-tab Trace Segment Attack}
In this section, we study the performance of \ST~on different types of segments in the single-tab trace if its front, tail or both ends is overlapping with other traces.
For all different percentages of segments in their single-tab trace, we focus on the impact of their types on the \ST~performance.
With the performance of \ST~on these segments, the importance of clean segment is in significance.
This is also the experience digit basis of the help of clean monitored segments to the \MT.

\keypoint{Setting. } 
In a partly-overlapping single-tab trace, there are seven types of trace segments: clean segment whose position is the front of trace, called as clean front segment and noted as Clean(Front), whose full trace is called as clean front full trace and noted as called as Full(Clean, Front); clean segment whose position is the tail of trace, called as clean tail segment and noted as  Clean(Tail), whose full trace is called as clean tail full trace and noted as  Full(Clean, Tail); clean segment whose position is the middle of trace, called as clean middle segment and noted as  Clean(Middle), whose full trace is called as clean middle full trace and noted as Full(Clean, Middle); overlapping segment whose position is the front of trace, called as overlapping front segment and noted as Overlap(Front); overlapping segment whose position is the tail of trace, called as overlapping tail segment and noted as Overlap(Tail).
On DS-19, with unmonitored traces, we overlapped its single monitored traces at 3 different positions: front, tail, and both ends.
We set 9 overlapping percentages from 0.1 to 0.9 with a step of 0.1.
For each type of segments with one fixed percentage, we set the training data set and test data set with the same data amount as the original DS-19, called as DS-19$_{overlap}$.
There are overall 27 different types of DS-19$_{overlap}$ according to their different three positions and different nine percentages.

\begin{figure}%[h!]
    \centering
    \includegraphics[scale=0.5]{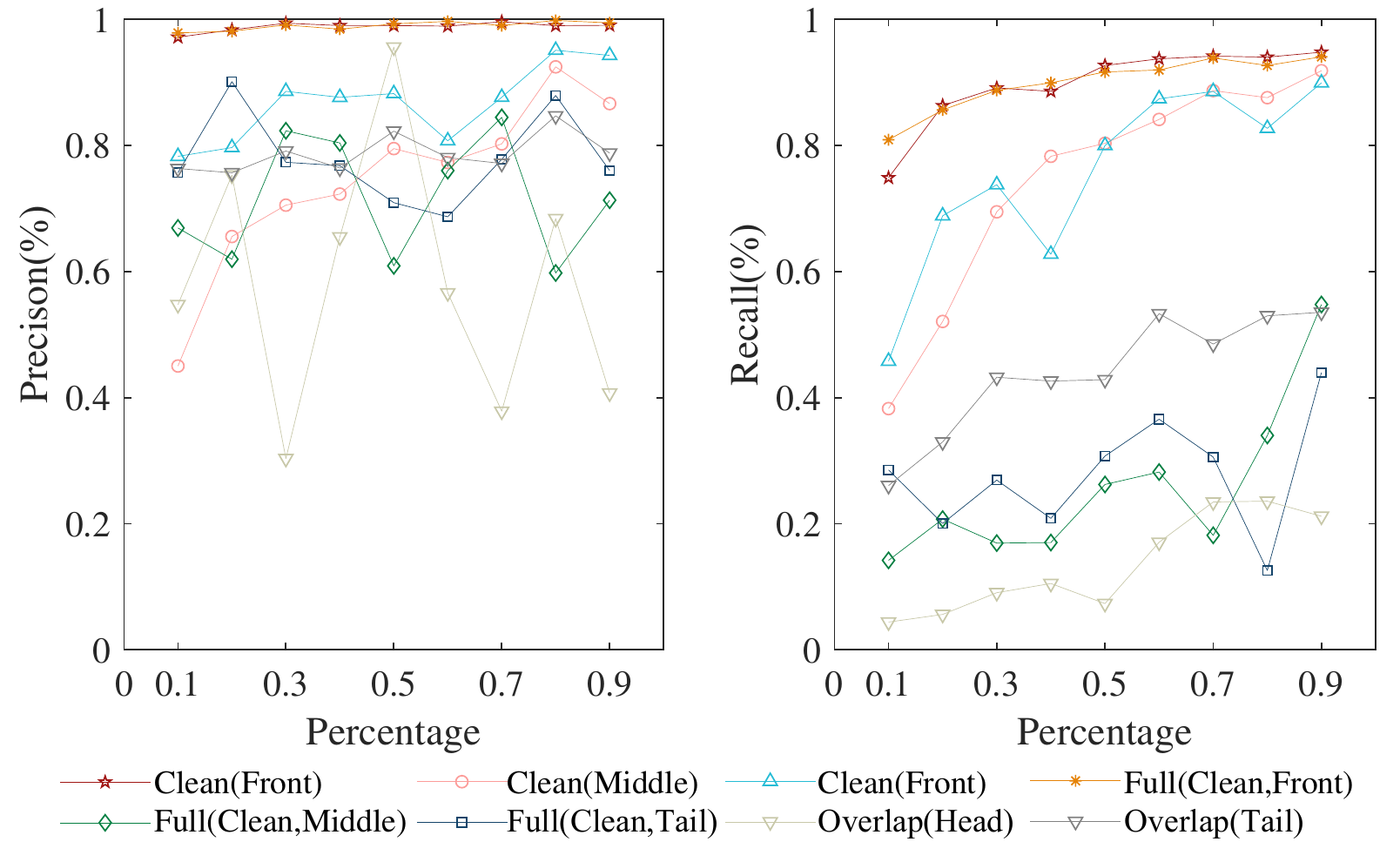}
    \caption{In the \emph{open-world} scenario, for different types of trace segments, the \WFA~performance of DF on the dataset DS-19$_{overlap}$.
    We plot evaluation results on the vertical Y-axis against percentages of segments in their full traces on the horizontal X-axis.
    Metrics: precision (\%) and recall (\%).
}
\label{fig:segment_ablation}
\end{figure}

\keypoint{Results}
The comparative results are reported in Fig \ref{fig:segment_ablation}.
We have the following observations and discussions.
\begin{enumerate}
\item
For the clean front segments, evaluated with precision metrics, the \ST~achieves the best performance on the clean front segment and its corresponding full trace, and the poorest performance on the overlapping front segments.
Similar results happen for the evaluation with the metrics of recall.
\item
For the clean tail segments, the attack performance on them is better than their corresponding full trace, only with a slight  advantage of precision but with a considerable  margin of recall.
\item
For the clean middle segments, the attack performance on them is also obviously better than their corresponding full trace, especially evaluating with the metrics of recall.
\item
For the overlapping segments, their performance is almost the poorest, even with a strong point of the overlapped tail segment on precision.
\item
For the position of segments, the front part is easiest to be attacked successfully.
If this part is overlapped, it would be most difficult to be attacked.
This is also the idea behind FRONT defense \cite{Gong2020}.
\item
For the types of a segment, the clean segment is undoubtedly the best choice for \ST.
This is why we think that the clean monitored segments would help our WF detection.
\end{enumerate}

\end{document}